\documentclass[useAMS,usenatbib]{mn2e}
\usepackage[dvips]{graphicx}
\usepackage{txfonts}
\usepackage{amssymb}
\usepackage{indentfirst}
\usepackage[T1]{fontenc}
\usepackage[latin1]{inputenc}

\begin{document}

\title[Soft band X/K luminosity ratios in late-type galaxies]{Soft band X/K luminosity ratios in late-type galaxies and constraints on the population of supersoft X-ray sources}

\author[\'A. Bogd\'an \& M. Gilfanov]{\'A. Bogd\'an$^{1}$\thanks{E-mail:
bogdan@mpa-garching.mpg.de (\'AB); gilfanov@mpa-garching.mpg.de (MG)} 
and
M. Gilfanov$^{1,2}$\footnotemark[1]\\
$^{1}$Max-Planck-Institut f\"ur Astrophysik, Karl-Schwarzschild-Str.1,
85741 Garching bei M\"unchen, Germany\\
$^{2}$Space Research Institute, Russian Academy of Sciences, Profsoyuznaya
84/32, 117997 Moscow, Russia}

\date{}

\maketitle

\begin{abstract}
We study X-ray to K-band luminosity ratios ($L_X/L_K$) of late-type galaxies in the $0.3-0.7$ keV energy range. From the \textit{Chandra} archive, we selected nine spiral and three irregular galaxies with point source detection sensitivity better than $ 5 \cdot 10^{36} \ \mathrm{erg \ s^{-1}} $ in order to minimize the contribution of unresolved X-ray binaries. In late-type galaxies  cold gas and dust may cause significant interstellar absorption, therefore we also demanded the existence of publicly available HI maps. The obtained $L_X/L_K$ ratios vary between $(5.4-68)\cdot10^{27} \mathrm{erg \ s^{-1} \ L_{K,\odot}^{-1}}$ exceeding by factor of $2-20$ the values obtained for gas-poor early-type galaxies. Based on these results  we constrain  the role of supersoft X-ray sources as progenitors of type Ia supernovae (SNe Ia). For majority of galaxies the upper limits range from $\sim 3\%$ to  $\sim 15\%$ of the SN Ia frequency inferred from K-band luminosity, but for a few of them  no meaningful constraints can be placed. On a more detailed level, we study individual structural components of spiral galaxies:  bulge and disk, and, for grand design spiral galaxies, arm and interarm regions.
\end{abstract}


\begin{table*}
\caption{The list of late-type galaxies studied in this paper.}
\begin{minipage}{18cm}
\renewcommand{\arraystretch}{1.3}
\centering
\begin{tabular}{c c c c c c c c c c}
\hline 
Name & Distance &  $L_{K} $               &  $N_{H}$ & Morphological &  SFR     & $ T_{\mathrm{obs}} $ & $ T_{\mathrm{filt}} $ & $L_{\mathrm{lim}}$            & $R$ \\ 
     & (Mpc)    &($\mathrm{L_{K,\odot}}$) &(cm$^{-2}$) & type        & ($\mathrm{M_{\odot} \ yr^{-1}}$)  & (ks)  & (ks)                  &  ($ \mathrm{erg \ s^{-1}} $)  & ($\arcsec$)   \\ 
     &   (1)    &            (2)           &   (3)      &     (4)     &      (5)                 &   (6) &  (7)                  &   (8)                         & (9)  \\
\hline 
M51     & $ 8.0^a $ &$ 6.4 \cdot 10^{10} $& $ 1.6 \cdot 10^{20} $ & SAbc       & 3.9 & $  90.9 $ & $  80.8 $ & $ 3 \cdot 10^{36} $ &  175 \\   
M74     & $ 7.3^a $ &$ 1.6 \cdot 10^{10} $& $ 4.8 \cdot 10^{20} $ & SA(s)c     & 1.1 & $ 104.4 $ & $  84.7 $ & $ 2 \cdot 10^{36} $ &  140 \\   
M81     & $ 3.6^b $ &$ 5.4 \cdot 10^{10} $& $ 4.1 \cdot 10^{20} $ & SA(s)ab    & 0.4 & $ 239.1 $ & $ 202.1 $ & $ 3 \cdot 10^{35} $ &  250 \\
M83     & $ 4.5^a $ &$ 3.9 \cdot 10^{10} $& $ 3.9 \cdot 10^{20} $ & SAB(s)c    & 2.8 & $  61.6 $ & $  52.1 $ & $ 1 \cdot 10^{36} $ &  200 \\
M94     & $ 4.7^a $ &$ 3.3 \cdot 10^{10} $& $ 1.4 \cdot 10^{20} $ & (R)SA(r)ab & 1.2 & $  76.9 $ & $  69.5 $ & $ 1 \cdot 10^{36} $ &  164 \\
M95     & $10.1^a $ &$ 3.3 \cdot 10^{10} $& $ 2.9 \cdot 10^{20} $ & SB(r)b     & 1.1 & $ 120.1 $ & $  99.6 $ & $ 3 \cdot 10^{36} $ &   88 \\
M101    & $ 7.4^a $ &$ 3.4 \cdot 10^{10} $& $ 1.2 \cdot 10^{20} $ & SAB(rs)cd  & 1.3 & $1070.6 $ & $ 833.3 $ & $ 2 \cdot 10^{35} $ &  167 \\
NGC2403 & $ 3.2^b $ &$ 5.0 \cdot 10^{9}  $& $ 4.2 \cdot 10^{20} $ & SAB(s)cd   & 0.4 & $ 224.0 $ & $ 184.2 $ & $ 2 \cdot 10^{35} $ &  140 \\
NGC3077 & $ 3.8^a $ &$ 2.6 \cdot 10^{9}  $& $ 4.0 \cdot 10^{20} $ & I0 pec     & 0.3 & $  54.1 $ & $  42.1 $ & $ 1 \cdot 10^{36} $ &  77.5 \\ 
NGC3184 & $11.1^c $ &$ 2.7 \cdot 10^{10} $& $ 1.1 \cdot 10^{20} $ & SAB(rs)cd  & 1.1 & $  66.7 $ & $  45.2 $ & $ 5 \cdot 10^{36} $ &  140 \\
NGC4214 & $ 2.9^a $ &$ 6.5 \cdot 10^{8}  $& $ 1.5 \cdot 10^{20} $ & IAB(s)m    & 0.2 & $  83.4 $ & $  53.6 $ & $ 4 \cdot 10^{35} $ &   70\\
NGC4449 & $ 4.2^a $ &$ 3.4 \cdot 10^{9}  $& $ 1.4 \cdot 10^{20} $ & IBm        & 0.4 & $ 102.1 $ & $  97.0 $ & $ 6 \cdot 10^{35} $ &   95\\
\hline \\
\end{tabular} 
\end{minipage}

\textit{Note.} Columns are as follows. (1) Distances. References are:  $^a$ \citet{karachentsev04}  -- $^b$ \citet{freedman01} -- $^c$ \citet{leonard02}. (2) Total near-infrared luminosity of the elliptic region described in column (9). (3) Galactic absorption \citep{dickey90}. (4) Taken from NED (http://nedwww.ipac.caltech.edu/). (5) Star-formation rate within the same ellipse. (6) and (7) Exposure times before and after data filtering. (8) Point source detection sensitivity in the $ 0.5-8 $ keV energy range. (9) Major axis of the studied elliptic regions. The orientation and shape of the regions were taken from K-band measurements (http://irsa.ipac.caltech.edu/applications/2MASS/). \\

\label{tab:list1}
\end{table*}

\section{Introduction}
SNe Ia are ``standardizable candles'' and hence can be used to determine the cosmological distance scale. Thus, they played a major role in establishing that the Universe is expanding at an accelerating rate, which fact directly pointed at the existence of the dark energy \citep{riess98,perlmutter99}. Despite the vast importance of SNe Ia, their progenitor systems are still debated.  

Recently we proposed an argument significantly  constraining  progenitor scenarios of SNe Ia in early-type galaxies \citep{nature}. We pointed out that the combined energy output of accreting and steady nuclear burning white dwarfs can be used to measure the rate at which white dwarfs increase their mass in galaxies. Thus, the contribution of the single degenerate scenario to the observed SN Ia frequency could be constrained. We concluded that in early-type galaxies no more than $\sim 5\%$ of SNe Ia are produced by white dwarfs accreting from a donor star in a binary system and detonating at the Chandrasekhar mass. 
Along similar (but not identical) line of arguments  \citet{distefano10}  compared  numbers of supersoft sources detected in several nearby elliptical and spiral galaxies with predictions of the single-degenerate scenario and found a discrepancy of about $ 2$ orders of magnitude.

It is tempting to apply the arguments of \citet{nature} to spiral and irregular galaxies. However,  it is important to realize that in case of  late-type galaxies  constraints on the X-ray output of accreting white dwarfs   do not necessary  lead to equally constraining upper limits on the contribution of the  single-degenerate scenario.  In young stellar environment  several  channels exist in which X-ray emission from growing white dwarfs may be significantly suppressed. These include high accretion rate configurations with optically-thick wind having large photospheric radius  \citep{hachisu96,hachisu99}, and white dwarfs accreting He-rich material, characterized by about an order of magnitude smaller nuclear energy release  \citep{iben94}. (Note that these channels are  not expected  to play significant role in much older elliptical galaxies considered by  \citet{nature}.)
Nevertheless, measurements of $L_X/L_K$ ratios are of much interest for late-type galaxies as well as they do constrain the populations of  supersoft X-ray sources and their possible role as progenitors of SNe Ia. This question is in the focus of the present paper.

There are several other important differences between late- and early-type galaxies relevant  to this study.
Firstly, it is the amount of cold interstellar gas and dust, absorbing the soft X-ray radiation from nuclear burning white dwarfs. Whereas it is negligible in elliptical galaxies, the enhanced and spatially variable absorption in younger galaxies strongly influences the observable X-ray emission from supersoft sources. This needs to be taken into account in comparing observations with predictions of SN Ia progenitor models. Secondly, it  is the presence and magnitude  of other X-ray emitting components in late-type galaxies, of which most important is ionized hot ISM. Whereas it is possible to find gas-free galaxies among ellipticals, all late-type galaxies appear to contain moderate to large amount of X-ray emitting gas. The exact distribution of this gas cannot be determined a priori, hence its contribution cannot be separated  and removed from other X-ray emitting components. In late-type galaxies high-mass X-ray binaries (HMXBs), located in star-forming regions, also can make a notable contribution \citep{grimm03}, and similarly to ellipticals, the population of low-mass X-ray binaries (LMXBs) is also present with luminosities down to $\sim10^{35} \ \mathrm{erg \ s^{-1}} $ \citep{gilfanov04}. Other type of faint unresolved sources, such as coronally active binaries, protostars, and young stars \citep{sazonov06} also contribute to the observed soft X-ray emission.  As all these components cannot be differentiated, we can  only give upper limits on the X-ray emission from the population of supersoft sources. Although same is true for early-type galaxies, the magnitude of contaminating factors is significantly larger in late-type galaxies.

The paper is structured as follows: in Sect. 2 we describe the sample selection and in Sect. 3 we introduce the analyzed data and discuss its reduction. The various X-ray emitting components are overviewed in Sect. 4 and the observed $L_X/L_K $ ratios are presented in Sect. 5. In Sect. 6 we investigate the role of supersoft sources as SN Ia progenitors.  We summarize in Sect. 7.

\section{Sample selection}
We aim to select a broad sample of nearby late-type galaxies, including spirals and irregulars. Precise knowledge of intrinsic absorption is critical for the comparison of observed X-ray luminosities with predictions of SN Ia models.  Therefore face-on galaxies are preferred due to the lack of projection effects. In a recent study \citet{walter08} studied the HI distribution of 34 nearby galaxies in full particulars. We used this sample as a starting point, as substantial fraction of their selected galaxies is well observed in X-ray wavelengths. The sample was filtered by demanding deep \textit{Chandra} observations, allowing to achieve a point source detection sensitivity of $ \lesssim 5\cdot 10^{36} \ \mathrm{erg \ s^{-1}}$ in the $0.5-8$ keV energy band. With this sensitivity we expect that the contribution of unresolved LMXBs and HMXBs do not influence significantly the observed $L_X/L_K$ ratios -- for quantitative analysis see Sect. \ref{sec:xbin}. There are $ 16 $ galaxies fulfilling these criteria. However, in case of four low-mass irregular galaxies (DDO53, DDO154, Holmberg I, IC2574) the total number of observed X-ray counts is too low to perform a detailed analysis, hence we excluded them from our sample. 

The final list of galaxies consists of nine spiral and three irregular galaxies, whose main properties are listed in Table \ref{tab:list1}. 

\section{Data reduction}
\subsection{\textit{Chandra} data}
We analyzed all publicly available observations of the selected galaxies which had an exposure time longer than $2$ ks. Totally, this yielded 71 observations (status: 06/2010). The overall exposure time of the data was $ T_{\mathrm{obs}} \approx 2.3 $ Msec. For each observation we extracted data of the S3 chip, except for Obs-ID 9553, where we used the entire ACIS-I array. The data was reduced with standard CIAO\footnote{http://cxc.harvard.edu/ciao} software package tools (CIAO version 4.1; CALDB version 4.1.3). 

The data reduction was performed similarly to that described in \citet{bogdan08}. First, we filtered the flare contaminated time intervals, after which the exposure time decreased by $ \lesssim 20\%$. The observed and filtered exposure times of the combined data are listed in Table \ref{tab:list1}. However, for point source detection the unfiltered data was used, since the longer exposure time outweighs the higher background periods. Observations for each galaxy were combined, they were projected in the coordinate system of the observation with the longest exposure time. To detect point sources we ran the CIAO wavdetect tool on the merged data separately in the $0.3-0.7$ keV and in the $0.5-8$ keV energy range, resulting source lists were combined. For the source detection algorithm we applied the same parameters that were discussed in \citet{bogdan08}. Accordingly, the obtained elliptic source regions contain large fraction, $\gtrsim 98\%$, of source counts. These output source cells were used to mask out point sources for further analysis of the unresolved emission. Note, that the source detection sensitivities ($L_{\mathrm{lim}}$), given in Table \ref{tab:list1}, refer to the $0.5-8$ keV band and are computed assuming $10$ net counts.

For galaxies significantly smaller than the field of view we estimated the background components using a combination of a number of regions away from galaxies. The angular size of four galaxies (M51, M74, M81, M83) exceeds or is comparable with the extent of the combined Chandra image, therefore we used the ACIS ``blank-sky'' files\footnote{http://cxc.harvard.edu/contrib/maxim/acisbg/} to estimate the background. As the instrumental background components of \textit{Chandra} vary with time, we renormalized the background counts using the $10-12$ keV count rates. To obtain point source detection sensitivities in the  $0.5-8$ keV energy range, we produced  exposure maps  assuming a power-law model with a slope of $\Gamma=2$. In the $0.3-0.7$ keV band we used exposure maps computed by assuming $N_H=10^{21} \ \mathrm{cm^{-2}}$ and a blackbody spectrum with $kT=100 $ eV.

\subsection{Infrared and HI data}
\label{sec:infrared}
The SN Ia rate in late-type galaxies can be either related directly to the mass of the galaxy \citep{mannucci05} with the scale depending on the galaxy type, or it can be decomposed into a mass related and a star-formation related components \citep{sullivan06}. Thus, we need to trace both the stellar mass and the star formation rate (SFR). The near-infrared data of the 2MASS Large Galaxy Atlas \citep{jarrett03} is known to be a good stellar mass tracer, hence we used K-band images for this purpose. The obtained 2MASS K-band images were background subtracted for all galaxies except for M51 and NGC3077, for these we estimated the background level by using several nearby regions off the galaxy.

The SFR was measured based on far-infrared images of the \textit{Spitzer Space Telescope}. After obtaining the luminosity at $70 \ \mathrm{\mu m}$ we converted it to total infrared luminosity \citep{bavouzet08}. This conversion incorporates notable systematic uncertainties whose measure depend on the employed far-infrared wavelength -- the least,  $19\%$, uncertainty applies for the $70 \ \mathrm{\mu m}$ data. Detailed discussion about the sources of uncertainties can be found in \citet{bavouzet08}. Based on the total infrared luminosities we computed the  corresponding star-formation rates using \citet{bell03}. The scatter associated with the obtained star-formation rate depends on the total infrared luminosity of the galaxy: for $10^{9} \ \mathrm{L_{\sun}} $ and $ 10^{11} \ \mathrm{L_{\sun}} $ it is $50\%$ and $20\%$, respectively \citep{bell03}. We note that no background is subtracted from  the \textit{Spitzer} $ 70 \ \mathrm{\mu m} $ images, provided by the \textit{Spitzer} archive, therefore we estimated it using nearby regions to the galaxies. 

Late-type galaxies contain a large amount of cold gas and dust, associated with the active star formation, causing notable interstellar absorption. As our goal is to study a rather soft X-ray band, it is crucial to map its distribution and measure the absorbing column accurately. For this purpose we used the HI maps of \citet{walter08} which give detailed images of the neutral hydrogen content of galaxies.

\section{Supersoft sources and other X-ray emitting components in late-type galaxies}

The goal of this investigation is to constrain the X-ray luminosity of nuclear burning white dwarfs. In order to maintain stable hydrogen burning on the surface of white dwarfs a rather high, $\dot{M} \gtrsim 10^{-7} \ \mathrm{M_{\sun}/yr}$, accretion rate is required \citep{nomoto07}. The energy of hydrogen fusion is released in the form of electromagnetic radiation, with bolometric luminosity of $L_{\mathrm{bol}} \sim10^{37}-10^{38} \ \mathrm{erg \ s^{-1}}$. The high bolometric luminosity means that some fraction of these sources may be detected by \textit{Chandra} as supersoft X-ray sources. However, because of  their low color temperature and  large absorption  in late-type galaxies, most of them will remain undetected by Chandra and will contribute to the unresolved soft emission. Obviously, in estimating the total X-ray output of nuclear burning white dwarfs, both resolved and unresolved supersoft sources need to be taken into account. However, there is a number of other X-ray emitting components contributing to unresolved emission in late-type galaxies, including genuinely  diffuse emission  from ionized ISM and various types of faint compact X-ray sources. These components play the role of  contaminating factors in measuring the X-ray luminosity of nuclear burning white dwarfs, and, as discussed below, account for the significant fraction of unresolved soft X-ray emission in late-type galaxies. As emission from these components cannot be separated from each other, it is only possible to obtain upper limits on the X-ray luminosity of nuclear burning white dwarfs. Their real luminosity is likely to be  smaller by a large (but unknown) factor.

\subsection{Resolved supersoft sources}

Bright supersoft sources were differentiated from other resolved point sources based on spectral properties. As the temperature of the hydrogen burning layer  is in the range of $\sim30-100$ eV, we conservatively included all resolved sources with hardness ratios corresponding to the blackbody temperature of $ kT_{bb} < 175 $ eV. Hardness ratios were computed based on three energy bands, $0.1-1.1$ keV as soft, $1.1-2$ keV as medium, and $2-7 $ keV as hard band, furthermore we applied the relations described by  \citet{distefano03} to identify supersoft sources.  Their total observed luminosities  are listed in Table \ref{tab:lxlk}.

\subsection{Hot ionized gas}
\begin{figure*}
\hbox{
\includegraphics[width=6cm]{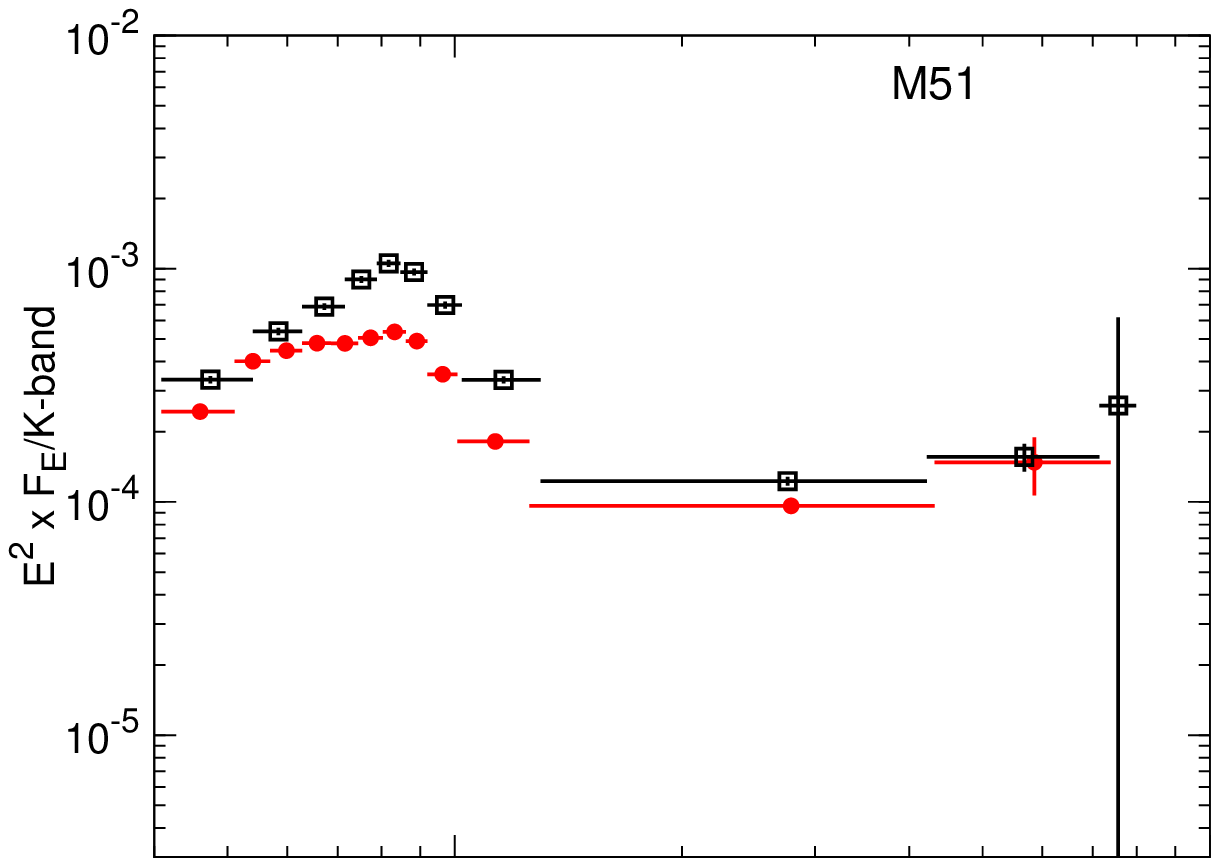}
\hspace{-0.3cm}
\includegraphics[width=6cm]{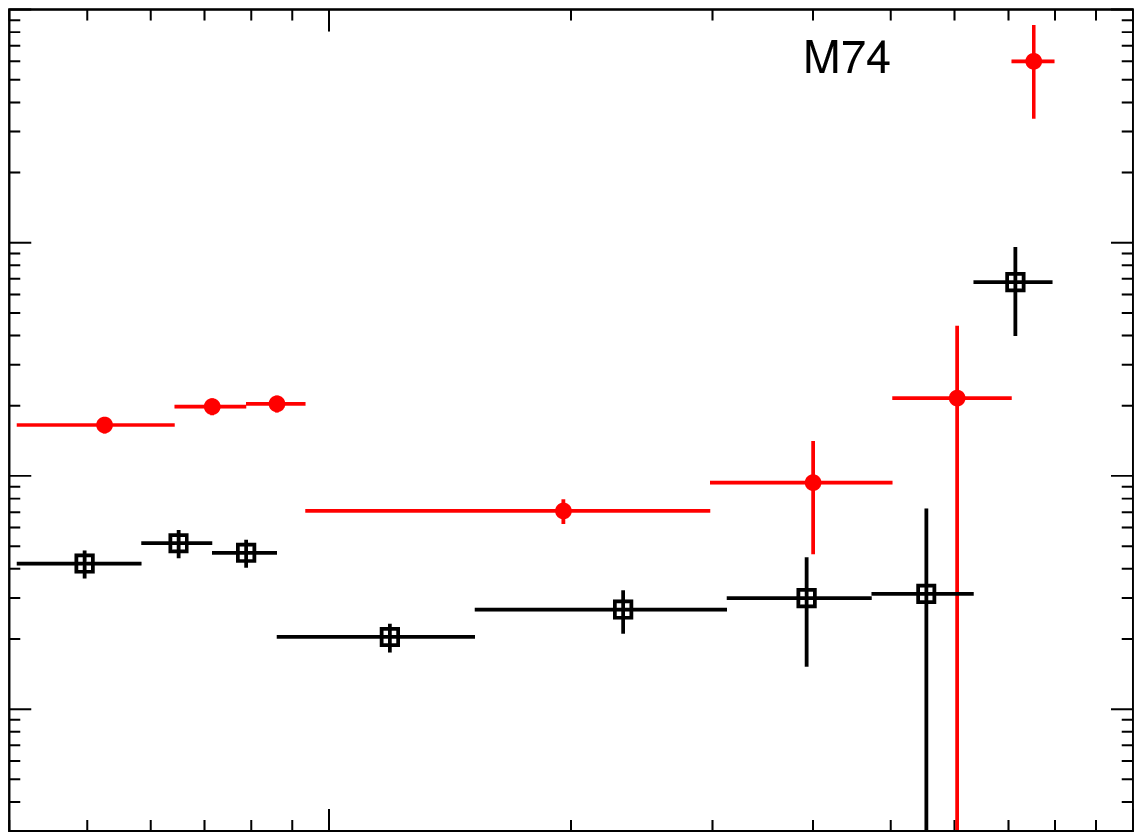}
\hspace{-0.3cm}
\includegraphics[width=6cm]{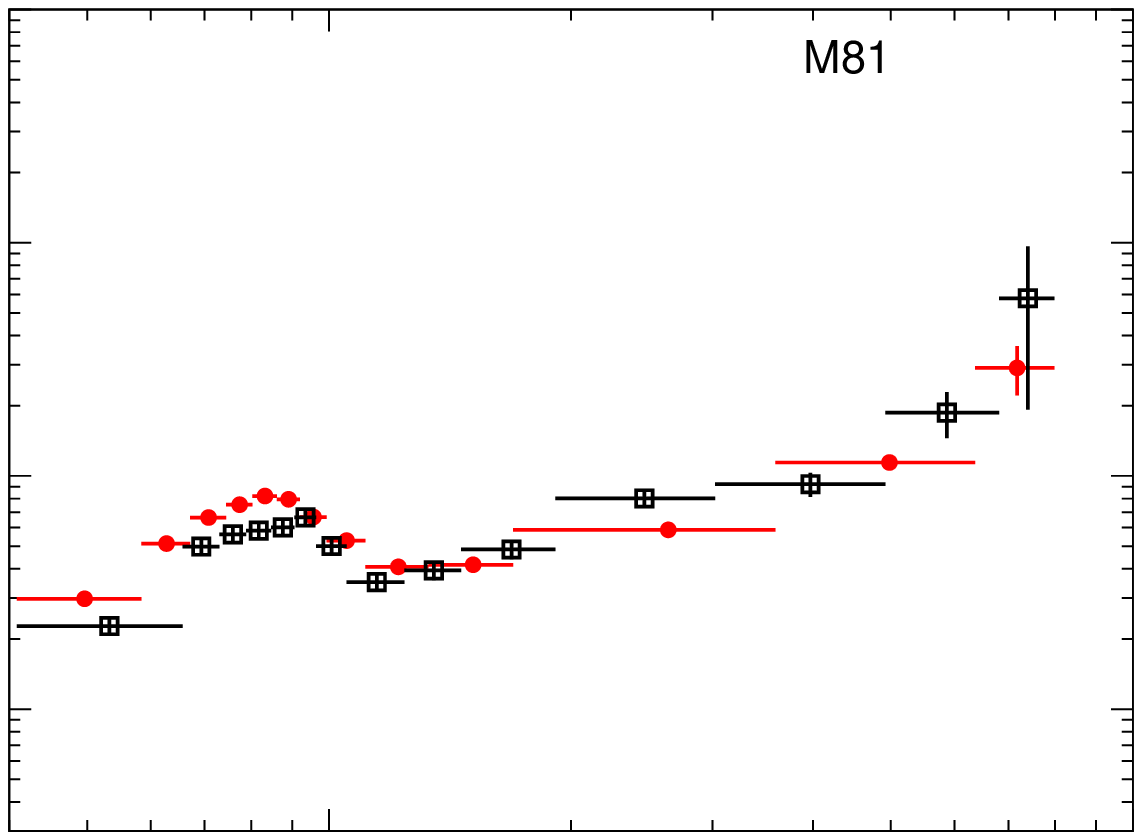}
}
\hbox{
\includegraphics[width=6cm]{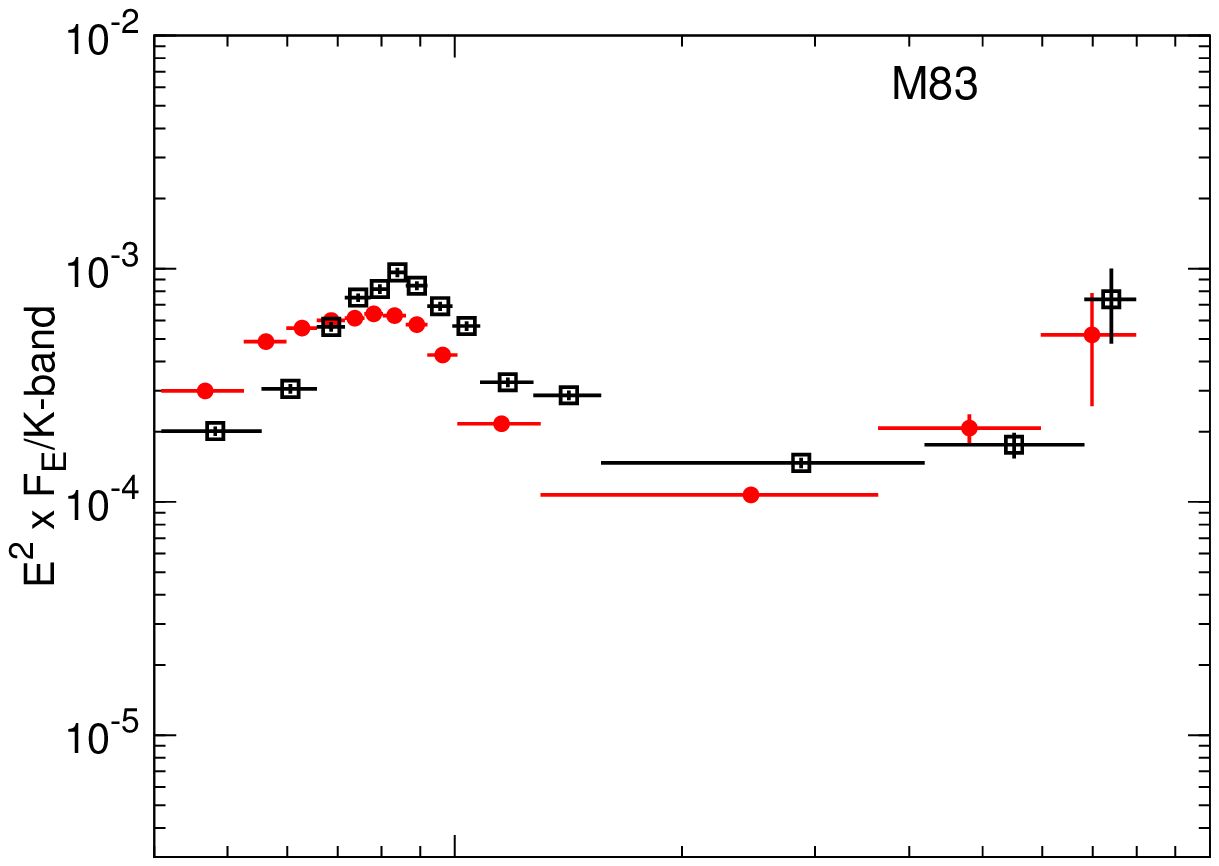}
\hspace{-0.3cm}
\includegraphics[width=6cm]{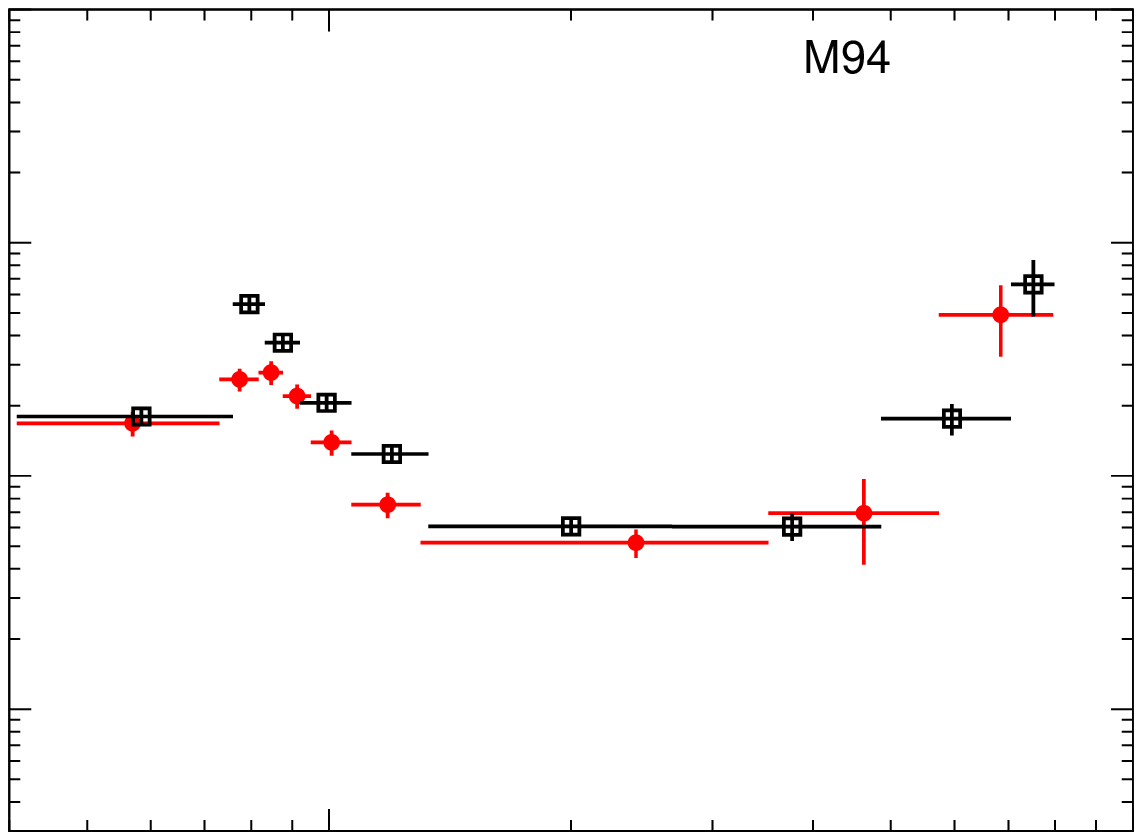}
\hspace{-0.3cm}
\includegraphics[width=6cm]{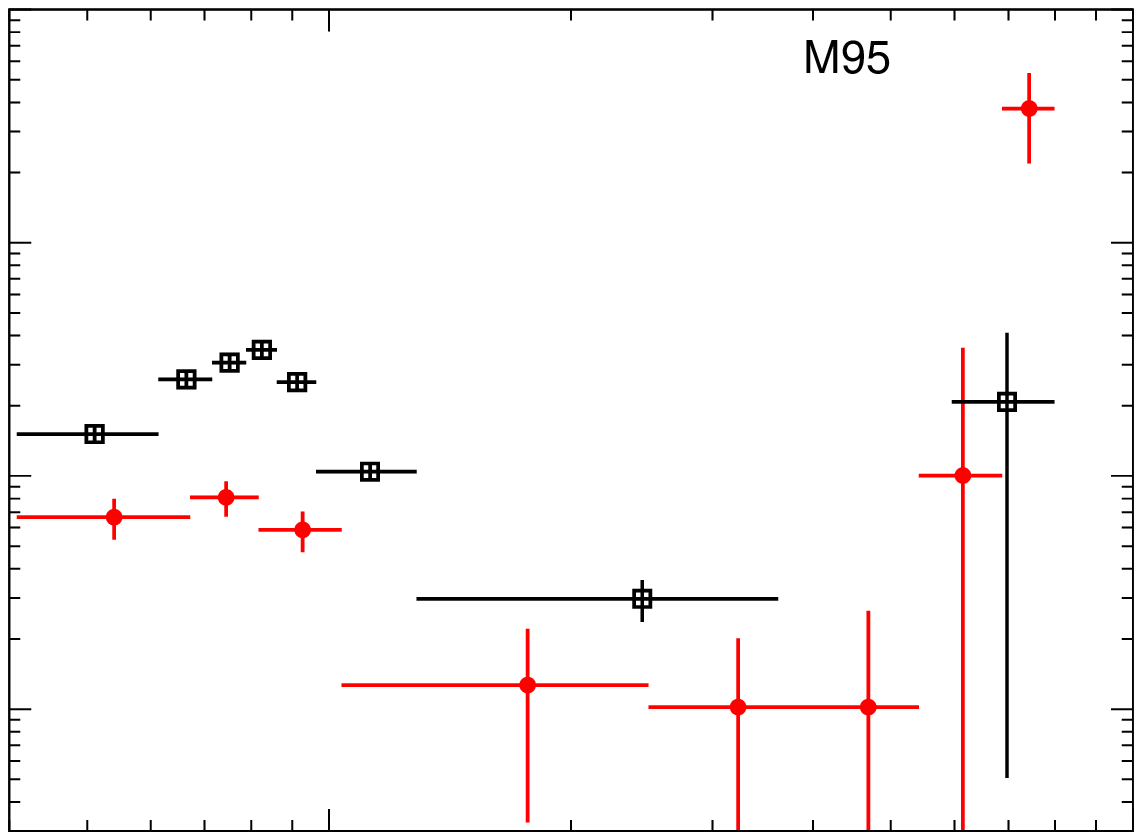}
}
\hbox{
\includegraphics[width=6cm]{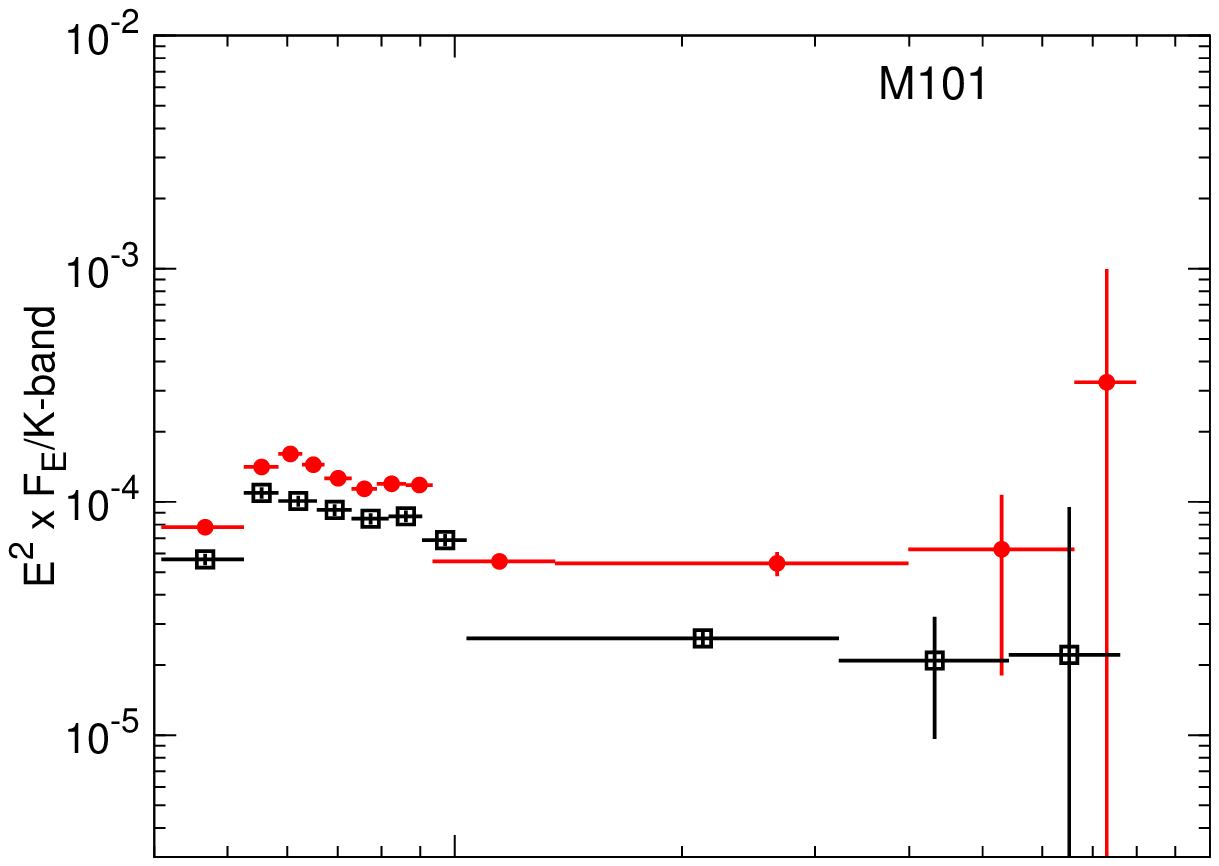}
\hspace{-0.3cm}
\includegraphics[width=6cm]{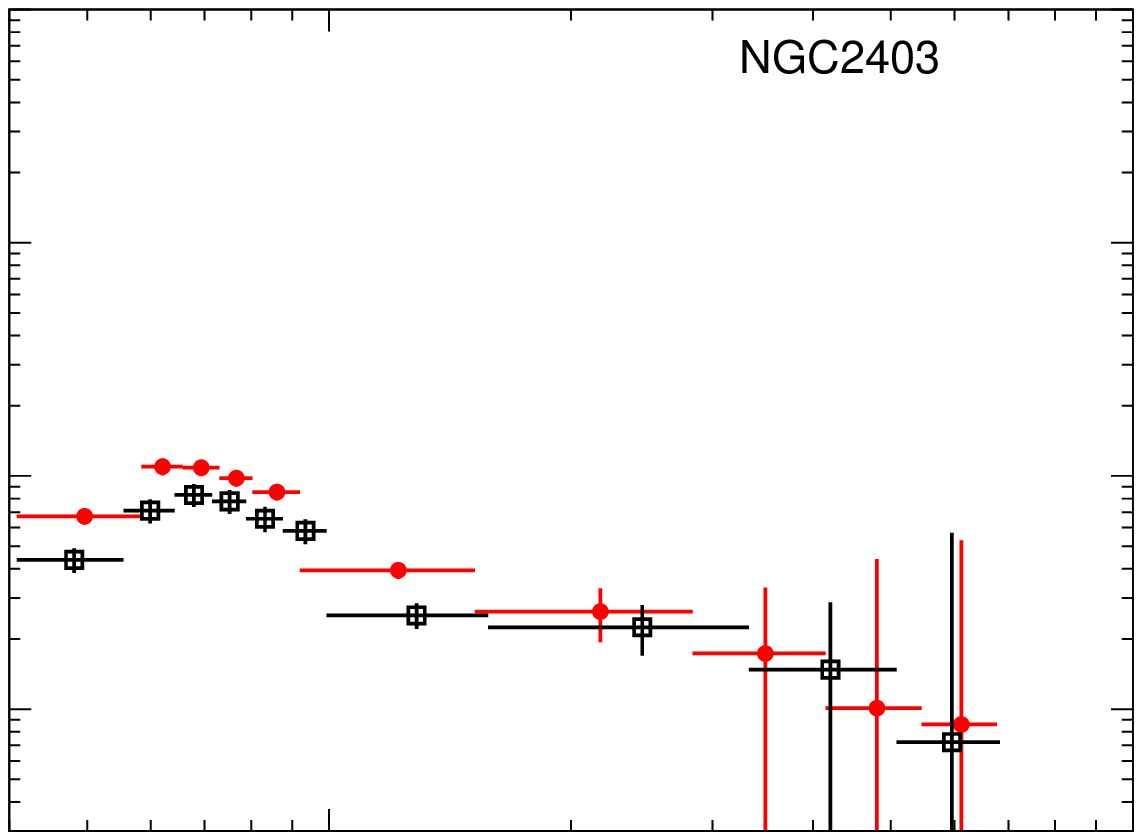}
\hspace{-0.3cm}
\includegraphics[width=6cm]{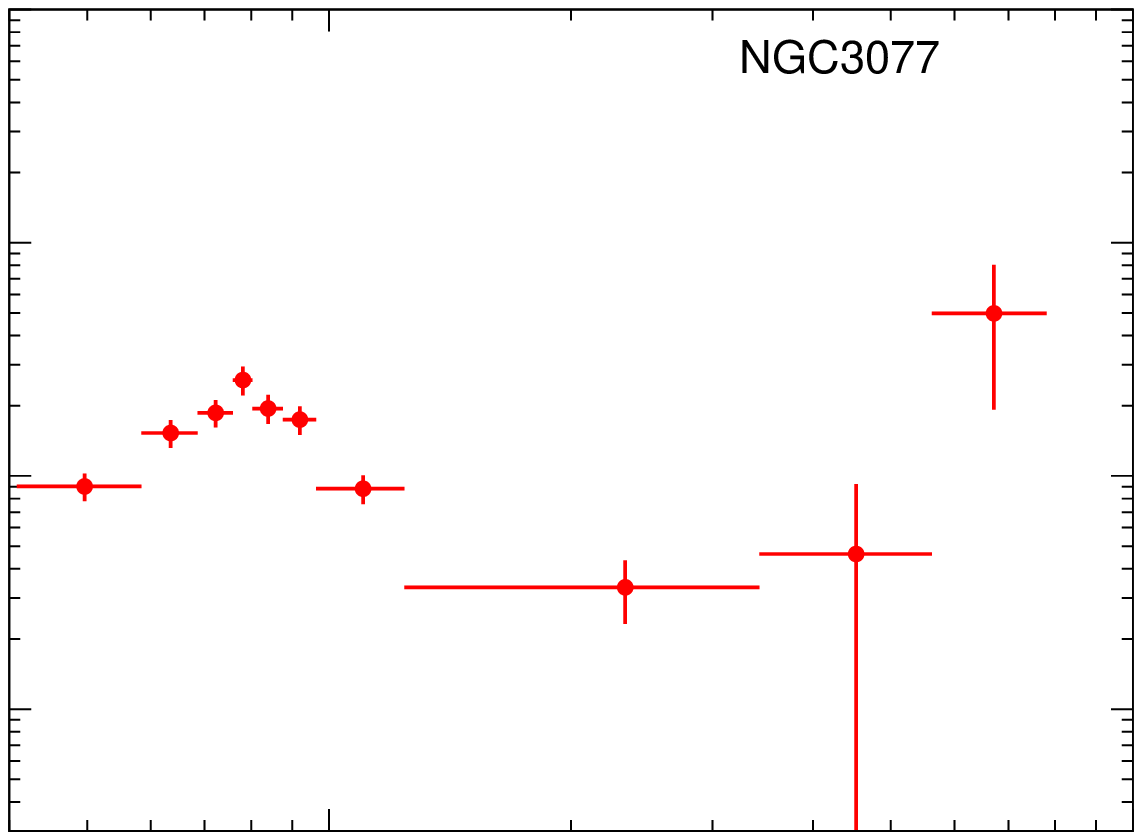}
}
\hbox{
\includegraphics[width=6cm]{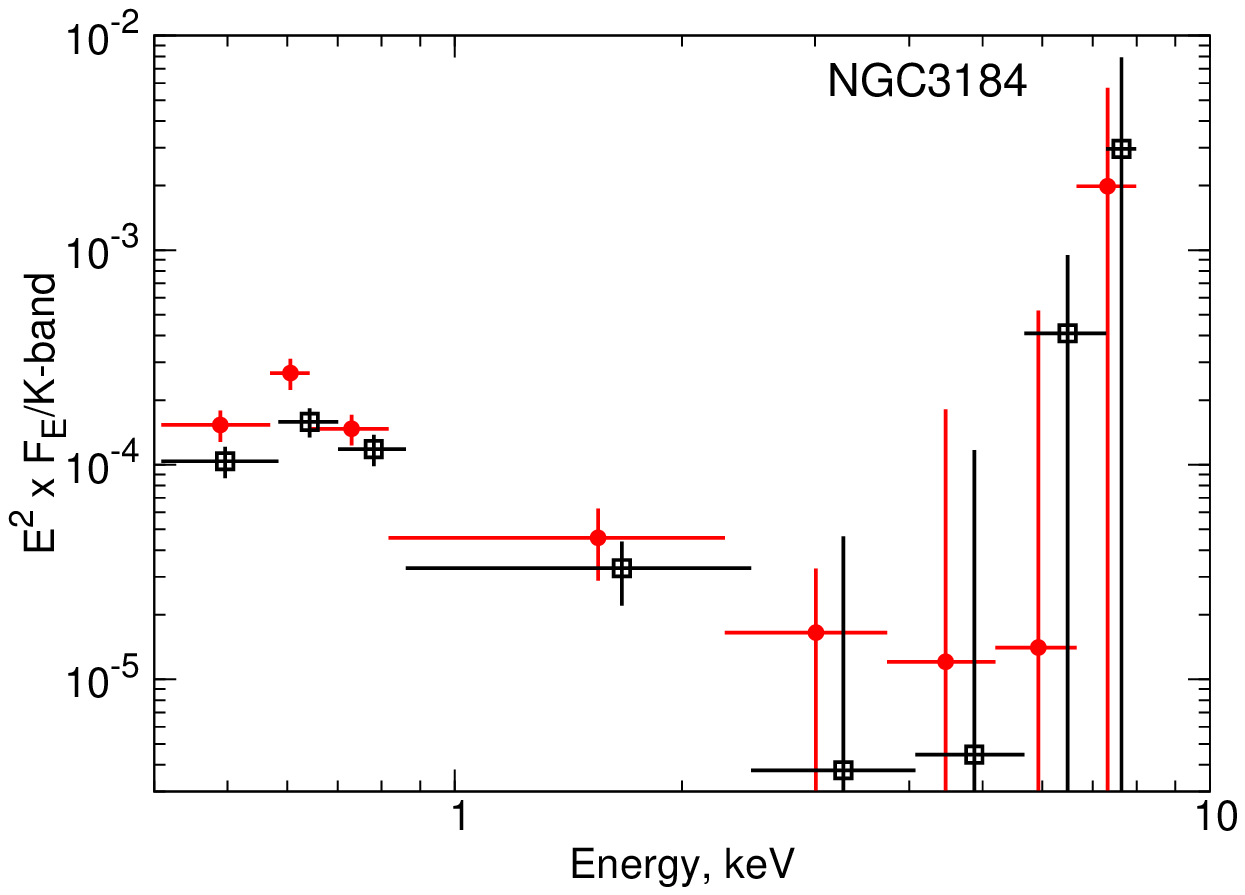}
\hspace{-0.3cm}
\includegraphics[width=6cm]{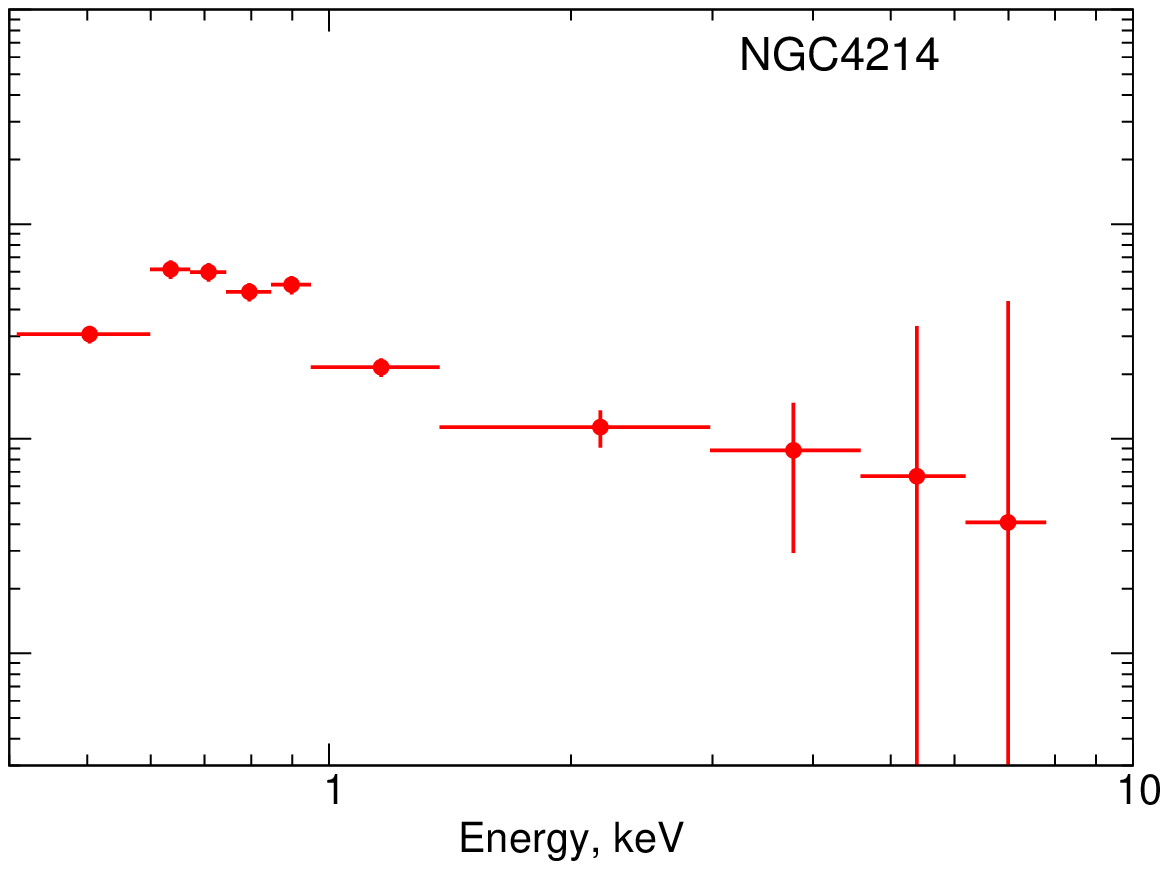}
\hspace{-0.3cm}
\includegraphics[width=6cm]{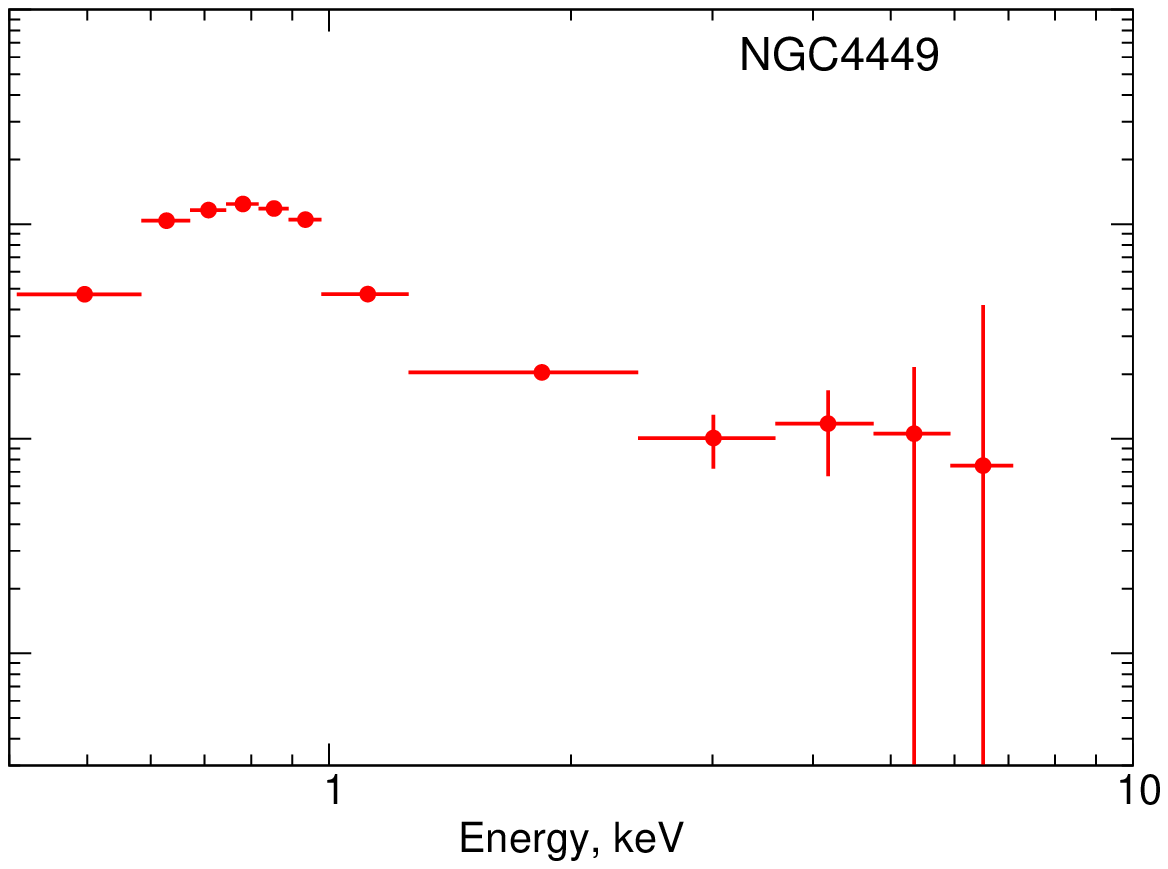}
}
\caption{Energy spectra of the galaxies in our sample. To facilitate comparison all spectra were normalized to the same level of K-band brightness and rescaled to the same distance of $10$ Mpc. All background components are subtracted. The scales on \textit{x}- and \textit{y}-axes are the same in all panels. For spiral galaxies, hollow squares (black) show the spectrum of the bulge, whereas filled circles (red) represent the spectrum of the disk. }  
\label{fig:enspec}
\end{figure*}

In late-type galaxies bright X-ray emission from hot ionized gas is expected. The total X-ray luminosity of this component may exceed the emission arising from unresolved compact X-ray sources. To reveal the presence of hot X-ray emitting gas we extracted spectra of the unresolved  emission in the studied galaxies, shown in Fig. \ref{fig:enspec}. For spirals we plot in the same panel the spectra of the bulge and the disk whereas for irregular galaxies the spectra refer to  the entire galaxy. To facilitate the comparison we normalized them to the K-band luminosity of $10^{11} \ \mathrm{L_{K,\odot}}$ and rescaled them to a distance of $10$ Mpc.

In each galaxy we detect a strong soft component which is presumably due to hot X-ray emitting gas, completely dominating X-ray emission  below $ \sim 1.5 $ keV. We performed spectral fits for each spectra and found that a two component model, consisting of an optically-thin thermal plasma emission model (MEKAL in XSPEC) and a power-law, gives an acceptable fit. As the main purpose of the present paper is to study the emission of unresolved supersoft sources, we do not study the results of spectral fits in details. Nevertheless, we mention that the best fit temperatures of the soft component are rather low, the obtained values are in the range of $ kT= 0.18-0.37 $ keV. These values are too high to be attributed to the population of supersoft sources, where the temperature of the hydrogen burning layer is $\sim 30-100 $ eV, hence this component originates most probably from hot ISM.  

It is obvious from Fig. \ref{fig:enspec} that emission from the hot ISM dominates luminosity in the $0.3-0.7$ keV  band.   It is also obvious that it cannot be separated from other  sources of soft X-ray emission. Therefore this is the main source of contamination in determining the luminosity of unresolved supersoft sources, much more significant than in the case of elliptical galaxies \citep{bogdan10}.

\begin{table*}
\caption{X-ray luminosities in the $ 0.3-0.7$ keV energy range of various X-ray emitting components in late-type galaxies.}
\begin{minipage}{18cm}
\renewcommand{\arraystretch}{1.3}
\centering
\begin{tabular}{c c c c c c c c c c}
\hline 
Name & $ N_{H,\mathrm{int}} $ & $L_{X,\mathrm{unres}} $ & $ L_{X,\mathrm{SSS}} $ & $ (L_X/L_K)_{\mathrm{total}} $ & $ (L_X/L_K)_{\mathrm{bulge}} $  & $ (L_X/L_K)_{\mathrm{disk}} $  & $ (L_X/L_K)_{\mathrm{arm}} $ & $ (L_X/L_K)_{\mathrm{interarm}} $  & $ (L_X/L_K)_{\mathrm{pred}} $\\ 
&($\mathrm{cm^{-2}}$) &($\mathrm{erg \ s^{-1}}$)&($\mathrm{erg \ s^{-1}}$)& ($\mathrm{erg \ s^{-1} \ L_{K,\odot}^{-1}}$)& ($\mathrm{erg \ s^{-1} \ L_{K,\odot}^{-1}}$) \\ 
     &   (1)  & (2)  &    (3)      &     (4)     &      (5)                 &   (6) &  (7)  & (8) & (9)                  \\
\hline 
M51     &  $ 7.7 \cdot 10^{20} $ & $ 2.7 \cdot 10^{39} $ & $ 3.2 \cdot 10^{38} $ & $ 4.7 \cdot 10^{28} $ & $ 4.7 \cdot 10^{28} $ & $ 4.5 \cdot 10^{28} $ &$ 3.9 \cdot 10^{28} $&$ 5.0 \cdot 10^{28} $& $ 3.1 \cdot 10^{29} $ \\
M74     &  $ 3.9 \cdot 10^{20} $ & $ 2.6 \cdot 10^{38} $ & $ 1.7 \cdot 10^{37} $ & $ 1.7 \cdot 10^{28} $ & $ 5.8 \cdot 10^{27} $& $ 2.3 \cdot 10^{28} $& $ 2.7 \cdot 10^{28} $& $ 2.1 \cdot 10^{28} $& $ 1.9 \cdot 10^{29} $ \\
M81     &  $ 7.8 \cdot 10^{20} $ & $ 1.4 \cdot 10^{38} $ & $ 1.5 \cdot 10^{38} $ & $ 5.4 \cdot 10^{27} $ & $ 6.4 \cdot 10^{27} $& $ 3.4 \cdot 10^{27} $& $ 3.2 \cdot 10^{27} $& $ 3.6 \cdot 10^{27} $& $ 1.7 \cdot 10^{29} $ \\
M83     &  $ 6.0 \cdot 10^{20} $ & $ 1.6 \cdot 10^{39} $ & $ 7.3 \cdot 10^{37} $ & $ 4.3 \cdot 10^{28} $ & $ 2.8 \cdot 10^{28} $& $ 5.2 \cdot 10^{28} $& $ 4.9 \cdot 10^{28} $& $ 5.5 \cdot 10^{28} $& $ 3.1 \cdot 10^{29} $ \\
M94     &  $ 4.6 \cdot 10^{20} $ & $ 2.9 \cdot 10^{38} $ & $ 2.3 \cdot 10^{37} $ & $ 9.5 \cdot 10^{27} $ & $ 1.1 \cdot 10^{28} $& $ 7.0 \cdot 10^{27} $&-- &-- &                                       $ 2.5 \cdot 10^{29} $ \\
M95     &  $ 2.1 \cdot 10^{20} $ & $ 3.9 \cdot 10^{38} $ & $ 2.0 \cdot 10^{37} $ & $ 1.2 \cdot 10^{28} $ & $ 1.9 \cdot 10^{28} $& $ 9.3 \cdot 10^{27} $&-- &-- &                                       $ 3.1 \cdot 10^{29} $ \\
M101    &  $ 1.0 \cdot 10^{21} $ & $ 4.3 \cdot 10^{38} $ & $ 4.3 \cdot 10^{37} $ & $ 1.4 \cdot 10^{28} $ & $ 9.8 \cdot 10^{27} $& $ 1.5 \cdot 10^{28} $&-- &-- &                                       $ 2.8 \cdot 10^{29} $ \\
NGC2403 &  $ 2.9 \cdot 10^{21} $ & $ 3.6 \cdot 10^{37} $ & $ 1.1 \cdot 10^{36} $ & $ 7.4 \cdot 10^{27} $ & $ 5.4 \cdot 10^{27} $& $ 7.8 \cdot 10^{27} $&-- &-- &                                       $ 1.4 \cdot 10^{28} $ \\
NGC3077 &  $ 8.5 \cdot 10^{20} $ & $ 3.2 \cdot 10^{37} $ & $ 3.5 \cdot 10^{36} $ & $ 1.4 \cdot 10^{28} $ & --&-- &-- &-- &                                                                             $ 3.3 \cdot 10^{29} $ \\
NGC3184 &  $ 2.7 \cdot 10^{20} $ & $ 6.5 \cdot 10^{38} $ & $ 4.3 \cdot 10^{36} $ & $ 2.4 \cdot 10^{28} $ & $ 1.4 \cdot 10^{28} $& $ 2.9 \cdot 10^{28} $&-- &-- &                                       $ 3.6 \cdot 10^{29} $ \\
NGC4214 &  $ 7.2 \cdot 10^{20} $ & $ 2.7 \cdot 10^{37} $ & $ 0 $                 & $ 4.2 \cdot 10^{28} $ & --&-- &-- &-- &                                                                             $ 7.3 \cdot 10^{29} $ \\
NGC4449 &  $ 3.0 \cdot 10^{21} $ & $ 2.2 \cdot 10^{38} $ & $ 9.5 \cdot 10^{36} $ & $ 6.8 \cdot 10^{28} $ & --&-- &-- &-- &                                                                             $ 3.4 \cdot 10^{28} $ \\
\hline \\
\end{tabular} 
\end{minipage}
\textit{Note.} Columns are as follows. (1) Average level of intrinsic $N_H$ (2) Total unresolved luminosity in the $0.3-0.7$ keV band. (3) Total luminosity of resolved supersoft sources in the $ 0.3-0.7 $ keV band. (4), (5), (6), (7), and (8) Observed $L_X/L_K$ ratios in the $0.3-0.7$ keV energy range for the entire galaxy, for its bulge, for its spiral arm regions, and for its interarm regions, respectively. The statistical errors vary between less than $1\%$ and $5\%$ (9)  Predicted $L_X/L_K$ ratios of supersoft sources in the single-degenerate scenario computed as described in the text; it is to be compared with column (4). 
\label{tab:lxlk}
\end{table*}

\subsection{Unresolved X-ray binaries}
\label{sec:xbin}
The total (including resolved sources)  X-ray emission of majority of galaxies is dominated by bright X-ray binaries. Their typical luminosity is in the range of $\sim10^{35}-10^{39}  \ \mathrm{erg \ s^{-1}}$, therefore X-ray binaries fainter than the detection threshold inevitably contribute to the luminosity of the unresolved emission. 

We estimate the contribution of LMXBs based on their luminosity function \citep{gilfanov04}, assuming that their average spectrum is described by a power-law model with a slope of $\Gamma=1.56$ \citep{irwin03}, and using a column density of $ 10^{21} \ \mathrm{cm^{-2}} $. Although the choice of $N_H$ is somewhat arbitrary, its particular value within the observed limits does not affect our conclusion. With these conditions the contribution of unresolved LMXBs to the observed $ L_X/L_K $ ratios is in the range of $ (6 \cdot 10^{24} - 1.1 \cdot 10^{26}) \ \mathrm{erg \ s^{-1} \ L_{K,\odot}^{-1}} $ in the $ 0.3-0.7 $ keV band, the upper limit is obviously associated with the worse source detection sensitivity. 

As active star formation is associated with disk of spirals and irregular galaxies, we also estimate the contribution of unresolved HMXBs to the $L_X/L_K $ ratios. We apply the luminosity function of \citet{grimm03} and the normalization according to \citet{shtykovskiy05} and assume an average power-law spectrum with slope of $ \Gamma=2 $ with a column density of $ N_{H} = 10^{21} \ \mathrm{cm^{-2}} $. Because the luminosity of HMXBs depends on the star-formation rate, their contribution to the $ L_X/L_K $ ratio will be highest in those galaxies where the star-formation rate per unit K-band luminosity ($\mathrm{SFR/L_K}$) is highest assuming the same point source detection sensitivity. However, this latter parameter varies in our sample, therefore we estimated the unresolved fraction of HMXB luminosity for each galaxy individually. The highest values were obtained for irregular galaxies, in these the unresolved HMXBs contribute with $ L_X/L_K  =  (4.1-7.6) \cdot 10^{26} \ \mathrm{erg \ s^{-1} \ L_{K,\odot}^{-1}}$, whereas in spirals we obtained $ L_X/L_K  =  (4.1 \cdot 10^{25} - 5.8 \cdot 10^{26}) \ \mathrm{erg \ s^{-1} \ L_{K,\odot}^{-1}}$. 

As we show in Sect. \ref{sec:numbers} the observed $ L_X/L_K  $ ratios exceed the estimated contribution of LMXBs and HMXBs by at least $ \sim 1-2 $ orders of magnitude. Thus, we can exclude with high confidence that the population of unresolved X-ray binaries make a major contribution to the derived $L_X/L_K$ ratios.

\subsection{Faint unresolved  sources}
Those nuclear burning white dwarfs which have low color temperatures ($kT\lesssim50$ eV) and/or are significantly absorbed will not be detected by \textit{Chandra} detectors as supersoft X-ray sources but will contribute to the unresolved emission. Besides these sources, however, there is a number of other types of X-ray sources, fainter than $ \lesssim 10^{35} \ \mathrm{erg \ s^{-1} } $, contributing to the unresolved emission. These include coronally active binaries, cataclysmic variables, protostars, young stars, supernova remnants etc. The populations of different classes of sources were studied in the Solar neighborhood by \citet{sazonov06} based on \textit{ROSAT} data. We used their specific luminosity value for the $ 0.1-2.4 $ keV range and estimated the $0.3-0.7$ keV emissivity for these sources  $L_X/L_K\sim {\rm few}\times 10^{27}\ \mathrm{erg \ s^{-1} \ L_{K,\odot}^{-1}}$. This can make a sizable contribution to the unresolved emission in few  galaxies with the lowest observed $ L_X/L_K $ ratios, but this contribution is not dominant.

\section{Results}
The X-ray luminosities of the unresolved emission ($L_{X,\mathrm{unres}} $) and resolved supersoft sources ($ L_{X,\mathrm{SSS}} $) are  presented in the first columns of Table \ref{tab:lxlk}. With the exception of M81, resolved supersoft sources contribute less than about $\sim 1-10\%$ to the total luminosity in the $0.3-0.7$ keV band. For M81,  the two luminosities are nearly equal. Its case will be further discussed below, but we also note here that the luminosity of resolved supersoft sources in this galaxy is dominated by one very bright source with $0.3-0.7$ keV luminosity of  $ 1.2 \cdot 10^{38}  \ \mathrm{erg \ s^{-1}}$.

\subsection{$L_X/L_K$ ratios}
\label{sec:numbers}

Further columns of Table \ref{tab:lxlk} present $ L_X/L_K $ ratios for galaxies and their different structural components. 
The average $L_X/L_K $ ratios range from $ (5.4 \cdot 10^{27}-6.8 \cdot 10^{28}) \ \mathrm{erg \ s^{-1} \ L_{K,\odot}^{-1} } $. Its lowest value  is observed in M81. Due to its  large angular extent,  the disk of this galaxy was mostly outside the Chandra field of view. This may explain, at  least in part, the smaller value of the $L_X/L_K $ ratio. The other end of the range is represented by the irregular galaxy  NGC4449. 

\begin{figure}
\hbox{
\includegraphics[width=8cm]{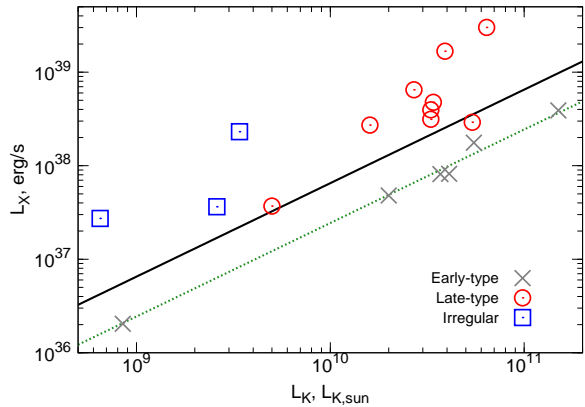}
}
\caption{The $0.3-0.7$ keV luminosity of unresolved emission and supersoft sources as a function of the K-band luminosity for early-type \citep{bogdan10}, spiral, and irregular galaxies. The solid line shows total emissivity of faint sources in the Solar    neighborhood from \citet{sazonov06}  recalculated to the $0.3-0.7$ keV band as described in the text. The dashed line   is the same but excluding contribution of  young stars. The Sagittarius  galaxy from \citet{bogdan10} falls outside the plot limits due to it its small K-band luminosity but lies on the extrapolation of the dashed curve, following the trend of early-type galaxies.}
\label{fig:lklx}
\end{figure}

Despite larger intrinsic absorption, the  $L_X/L_K$ ratios exceed by factor of about $  2-20 $ those of early-type galaxies \citep{bogdan10}. This is further illustrated by Fig. \ref{fig:lklx} where X-ray  luminosities are plotted against K-band luminosities. The early-type galaxy sample is taken from \citet{bogdan10}. Although the contribution of bright resolved supersoft sources is included, it does not exceed $\sim 10\%$ (Table \ref{tab:lxlk}). Therefore this plot shows, essentially, the luminosity of unresolved emission.  
As in the case of early-type galaxies the  contribution of unresolved X-ray binaries may be of some importance,  we used X-ray luminosities transformed to the uniform source detection sensitivity of $2 \cdot 10^{36} \ \mathrm{erg \ s^{-1}}$. The straight lines show the emissivity of faint compact sources in the Solar neighborhood from \citet{sazonov06}, including (solid) and excluding (dashed) the contribution of young stars.
To convert their $L_X/L_K$ ratio from the $0.1-2.4$ keV to the $0.3-0.7$ keV band we assumed $ N_H =10^{21} \ \mathrm{cm^{-2}} $ and an optically-thin thermal plasma emission model (MEKAL) with $kT=0.3$ keV; if a $\Gamma=2$ power-law is assumed,  the lines move downwards by a factor of $\sim 2$. The agreement of soft X-ray luminosity of gas-poor early type galaxies with the old population of the Solar neighborhood is remarkable (we note that due to small investigated volume,  the latter does not include any supersoft sources, however, as mentioned above, their contribution in the soft luminosity of early-type galaxies does not exceed $\sim 10-20\%$). Late-type galaxies, on the other hand, lie, with a large scatter, significantly above the line, corresponding to the Solar neighborhood. This is explained by contribution of the hot ionized ISM.

To further investigate spiral galaxies  we measured $ L_X/L_K $ ratios  for their bulges and disks. The separation was made  based on near-infrared images. We did not find very large difference -- the $ L_X/L_K $ ratios typically  do not differ by more than a factor of $\sim 2$, except for M74 where the disk value is larger by a factor of $\sim4$. There is a number of factors which may affect $ L_X/L_K $ ratios: the younger stellar population of the disk will tend to have higher $L_X/L_K$, whereas the more centrally concentrated hot X-ray emitting gas and the lower intrinsic absorption will  increase   its value in the bulge. Combination of these -- and possibly other -- effects may be responsible for observed scatter in $L_X/L_K$ values. 

Four galaxies in our sample (M51, M74 , M81, and M83) have grand design spiral arms allowing us to measure $L_X/L_K$ ratios of spiral arms and interarm regions separately, obtained values are given in Table \ref{tab:lxlk}. The arm and interarm regions were separated based on optical and near-infrared images. To our surprise, we did not find large difference between spirals arms and interarm regions, their  $L_X/L_K$ ratios being different by no more than  $\sim30\%$. The fact that the $L_X/L_K $ ratios in spiral arms are not suppressed by the significantly higher intrinsic absorption may suggest that supersoft sources do not contribute significantly to the soft  X-ray emission form the disks of spiral galaxies, but   X-ray sources  with  harder spectrum dominate.  The most likely candidate for such sources is  the emission from ISM (Fig. \ref{fig:enspec}).

\begin{figure*}
\hbox{
\includegraphics[width=6cm]{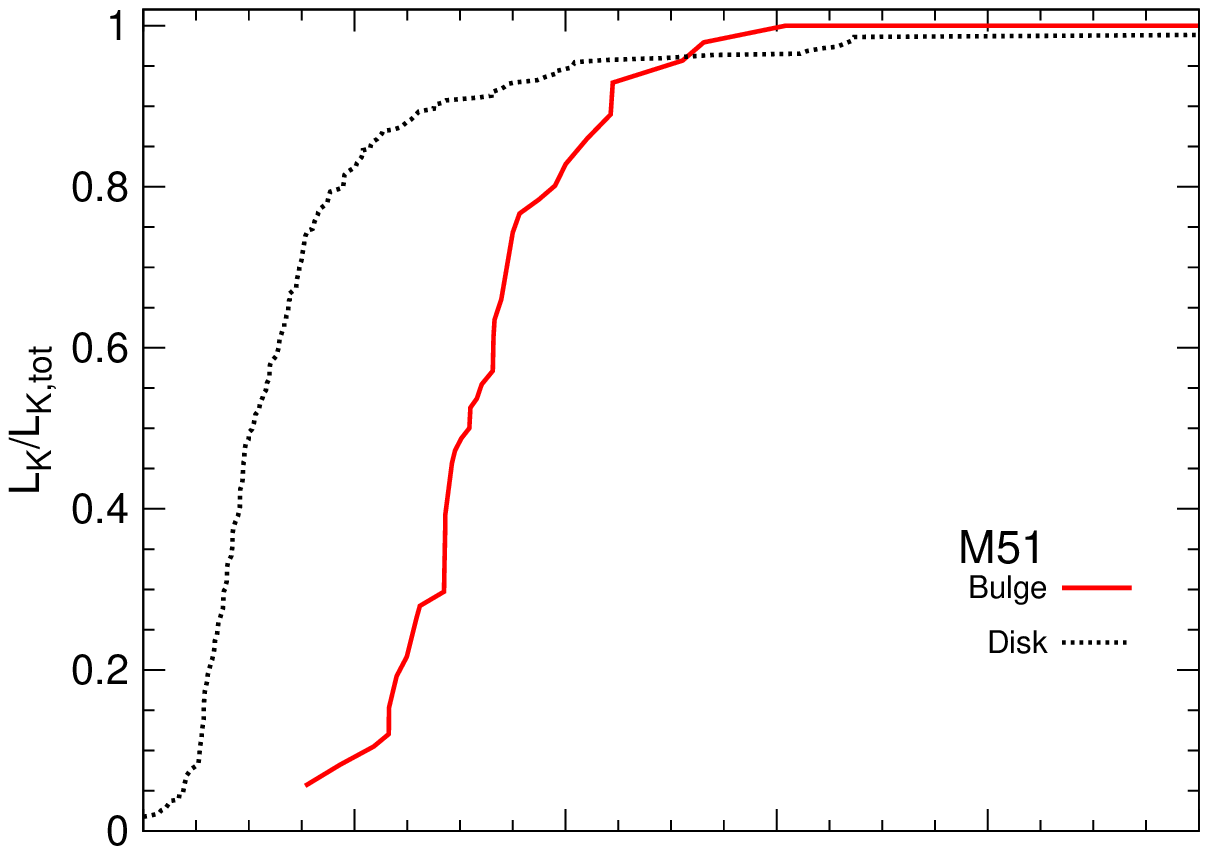}
\hspace{-0.3cm}
\includegraphics[width=6cm]{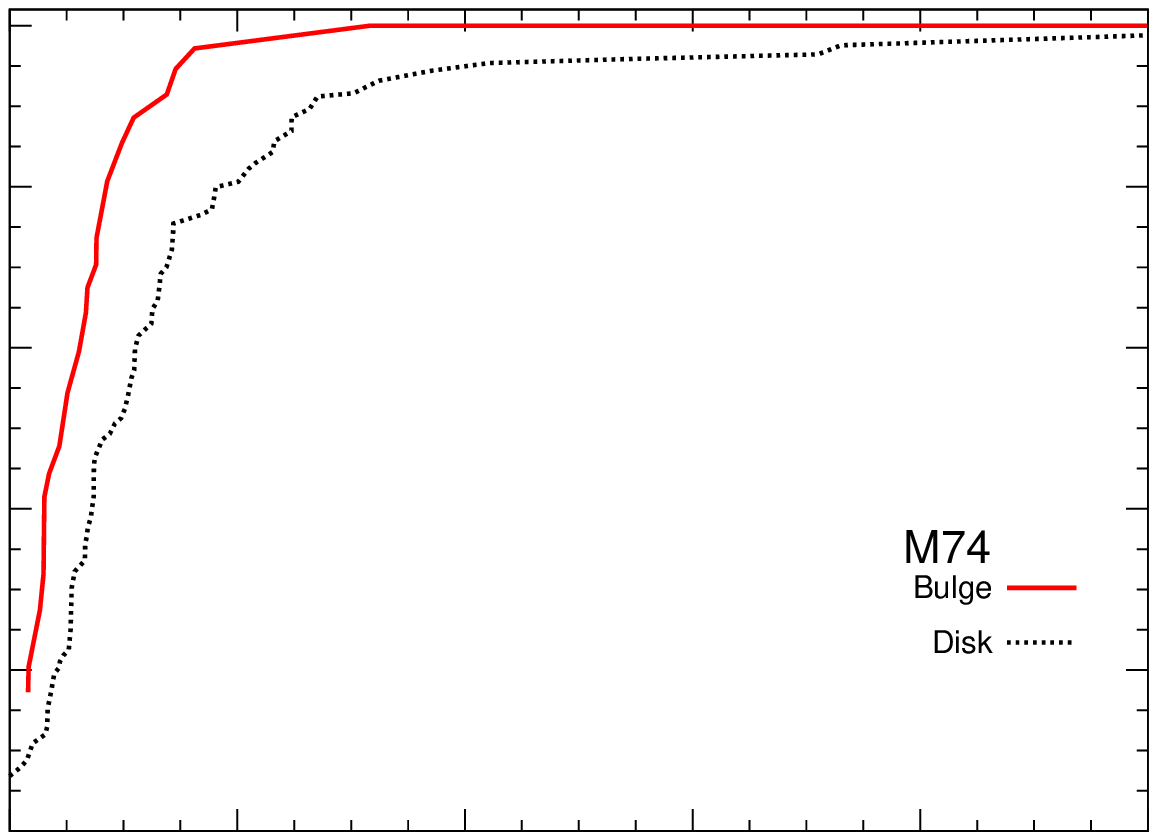}
\hspace{-0.3cm}
\includegraphics[width=6cm]{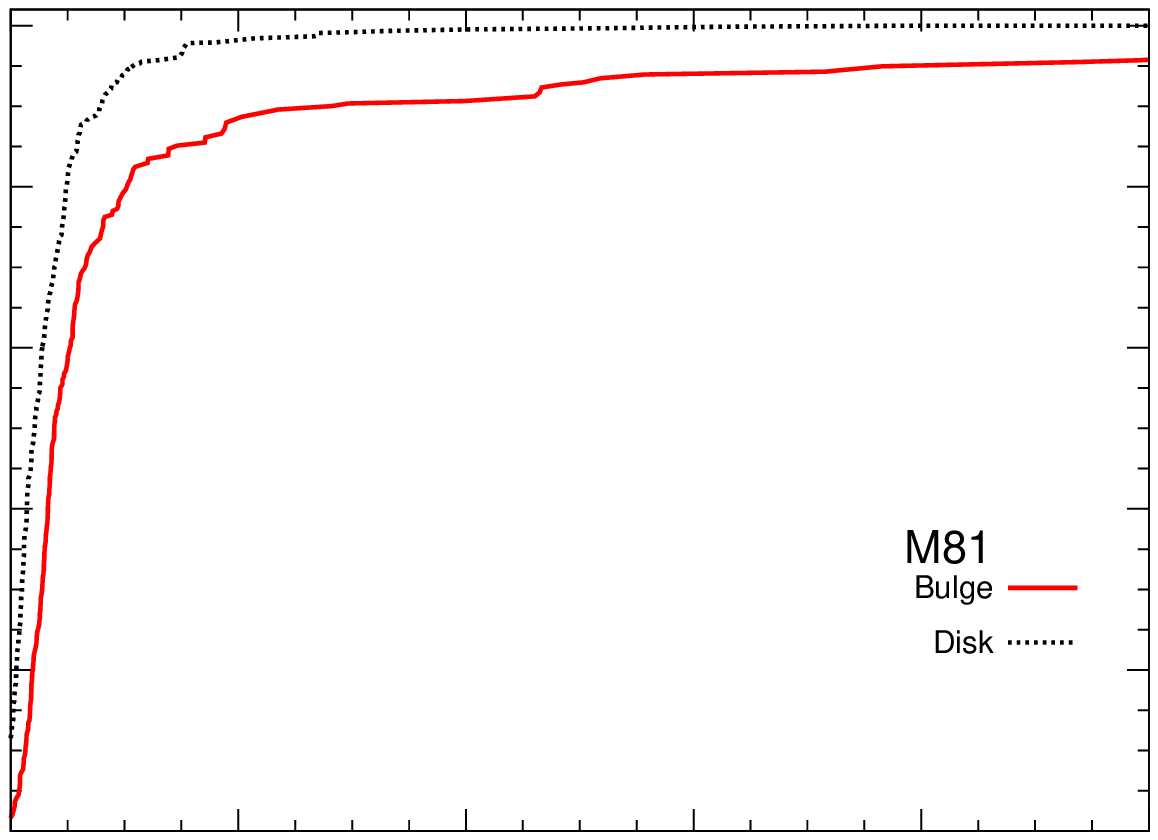}
}
\hbox{
\includegraphics[width=6cm]{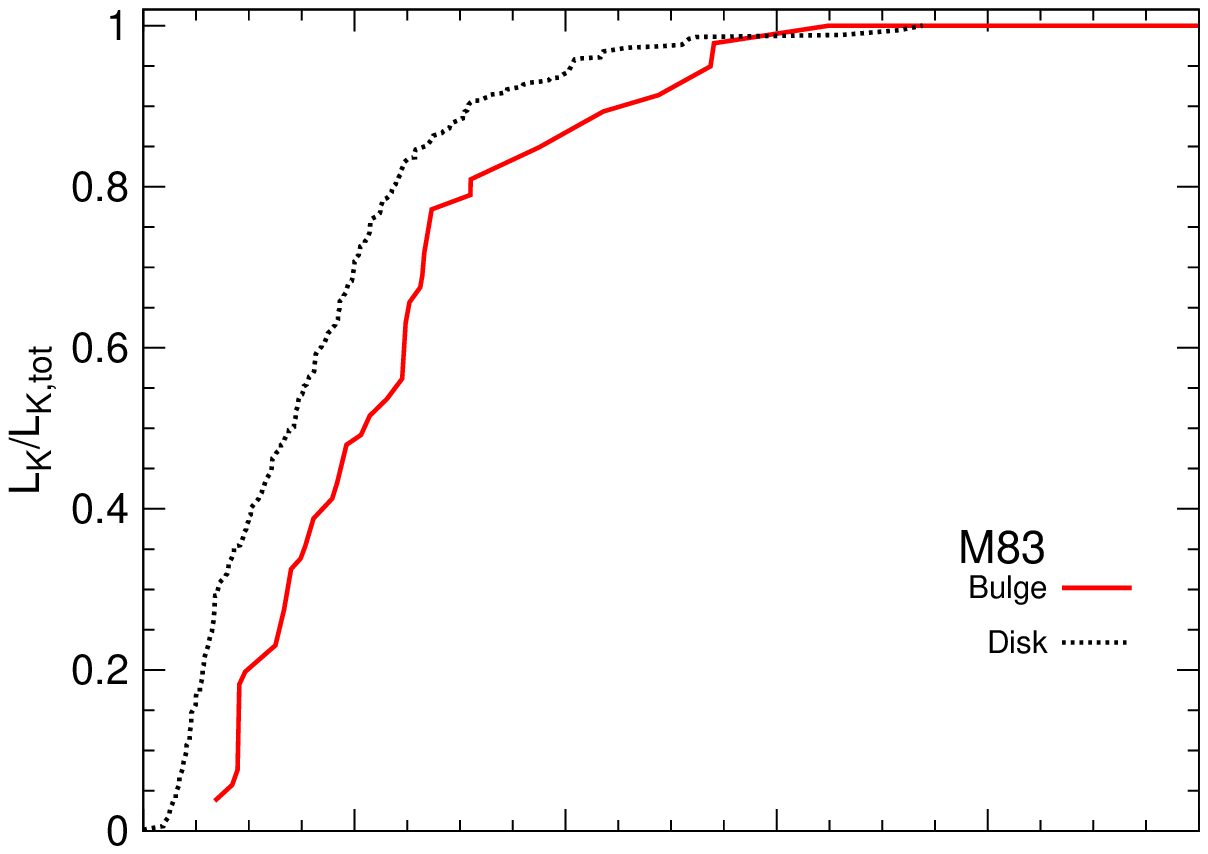}
\hspace{-0.3cm}
\includegraphics[width=6cm]{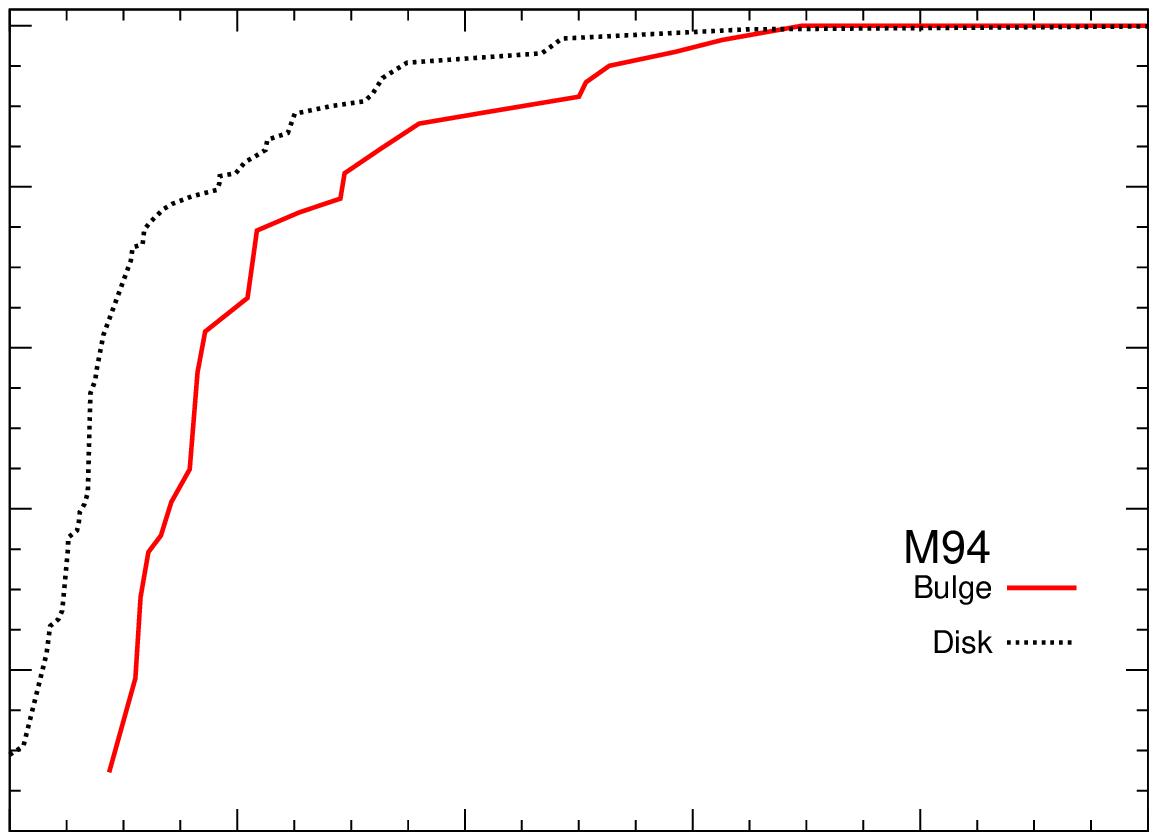}
\hspace{-0.3cm}
\includegraphics[width=6cm]{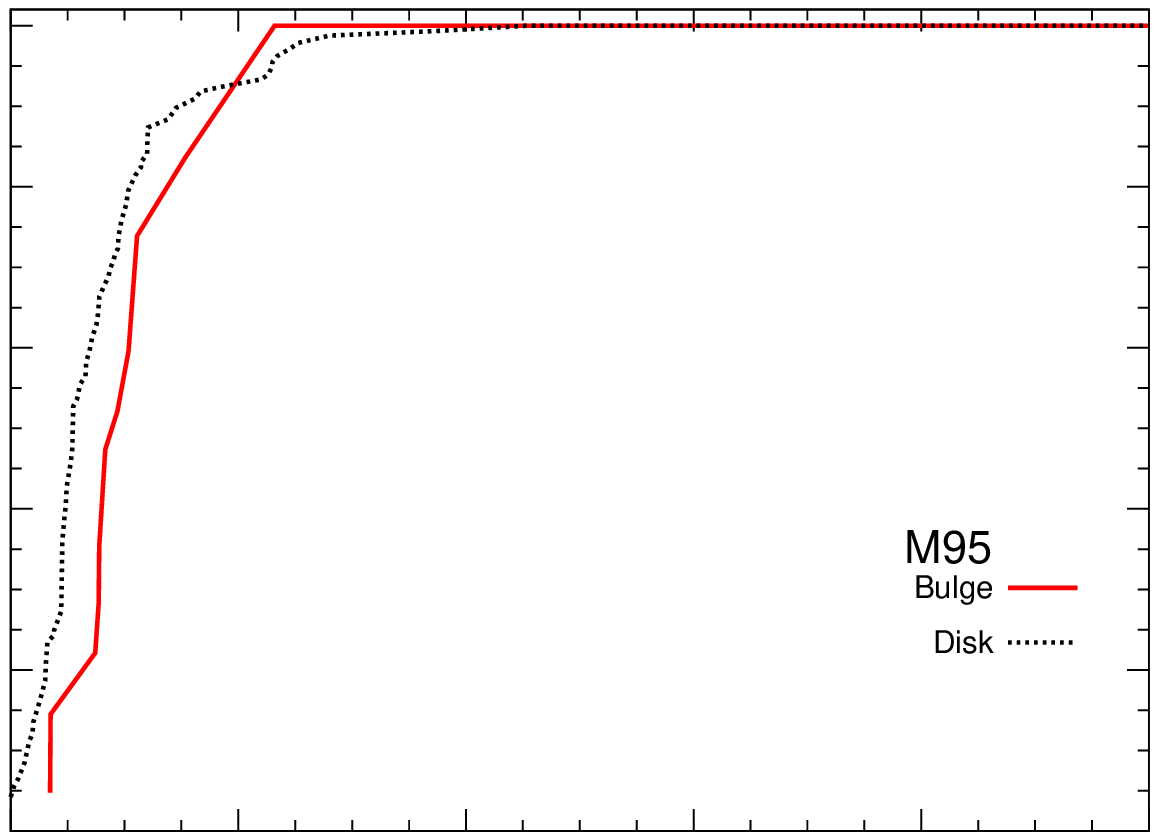}
}
\hbox{
\includegraphics[width=6cm]{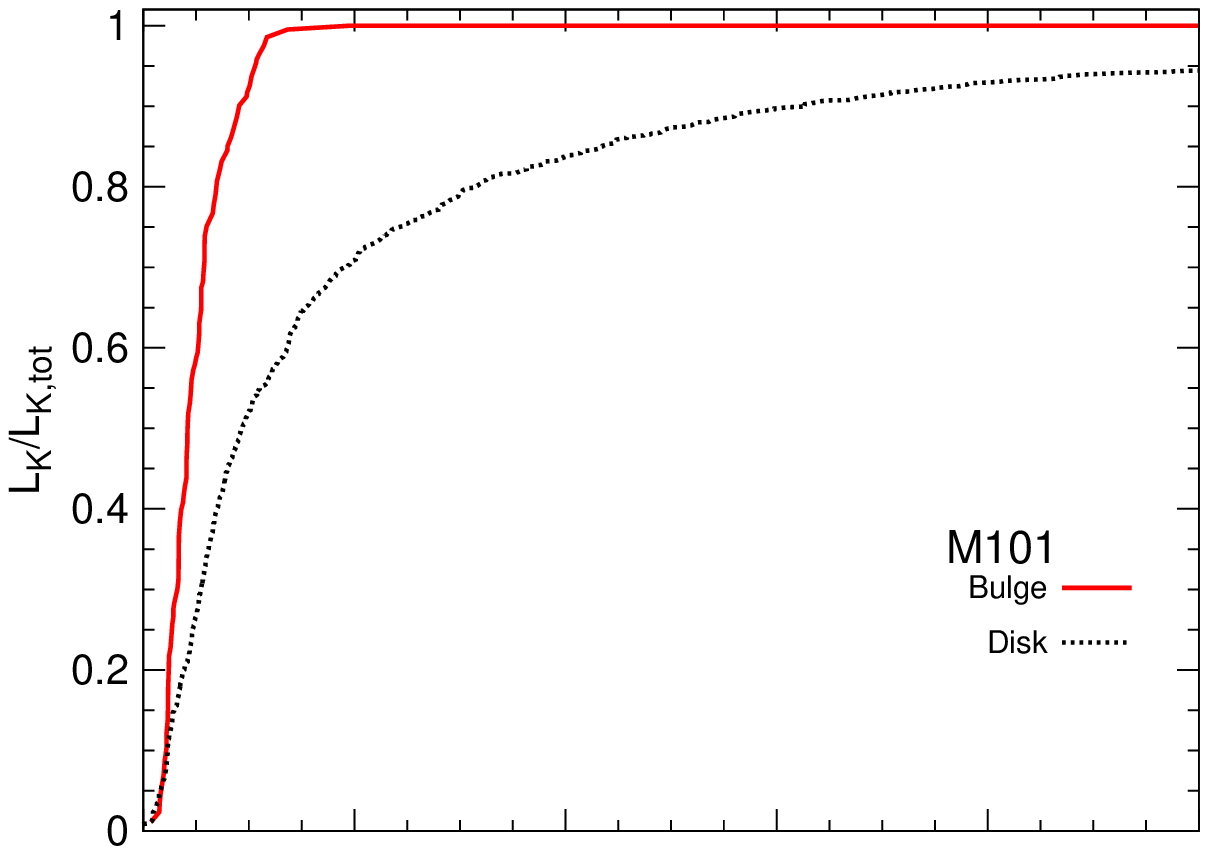}
\hspace{-0.3cm}
\includegraphics[width=6cm]{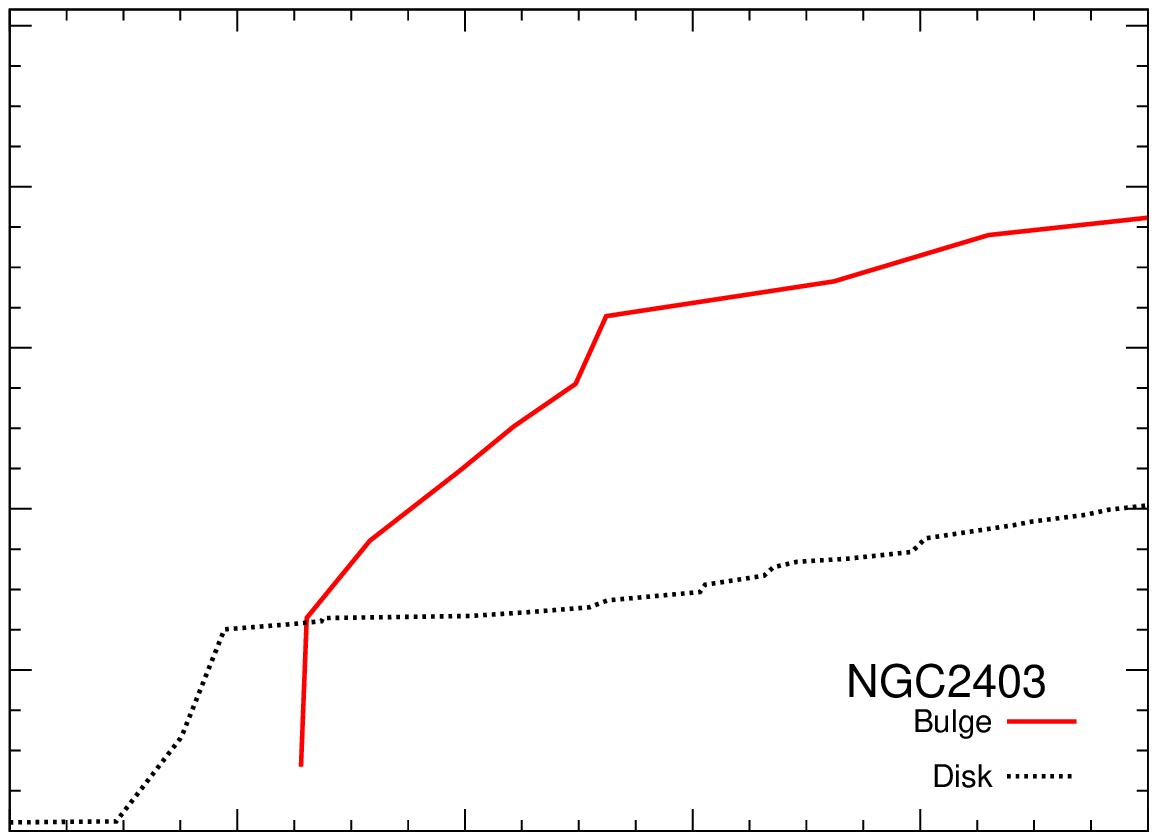}
\hspace{-0.3cm}
\includegraphics[width=6cm]{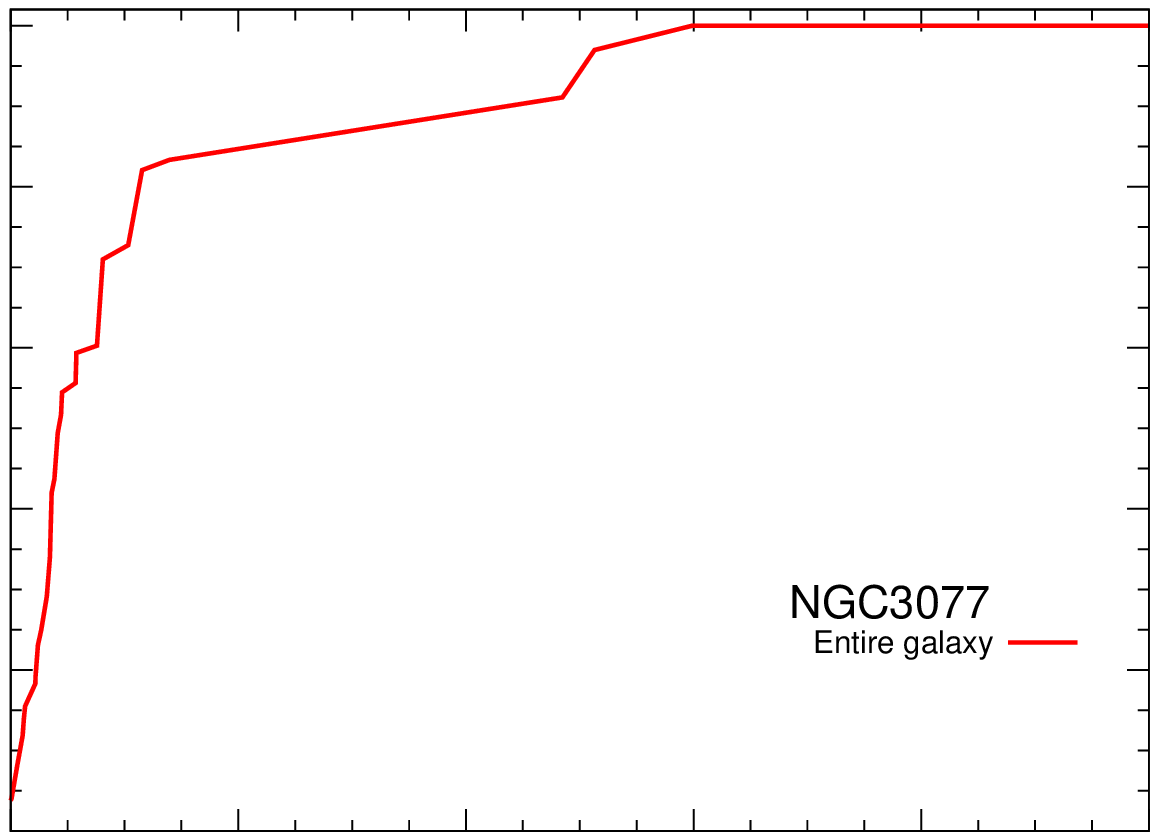}
}
\hbox{
\includegraphics[width=6cm]{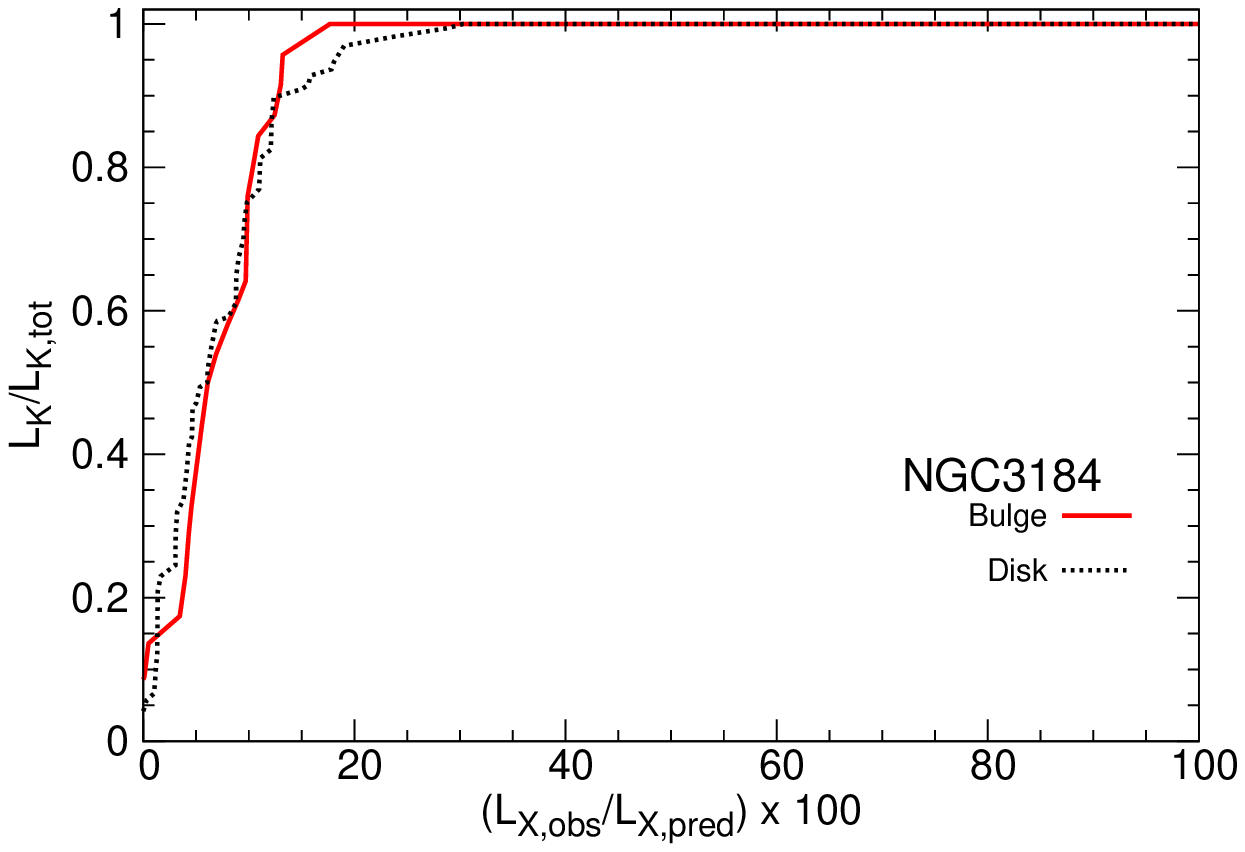}
\hspace{-0.3cm}
\includegraphics[width=6cm]{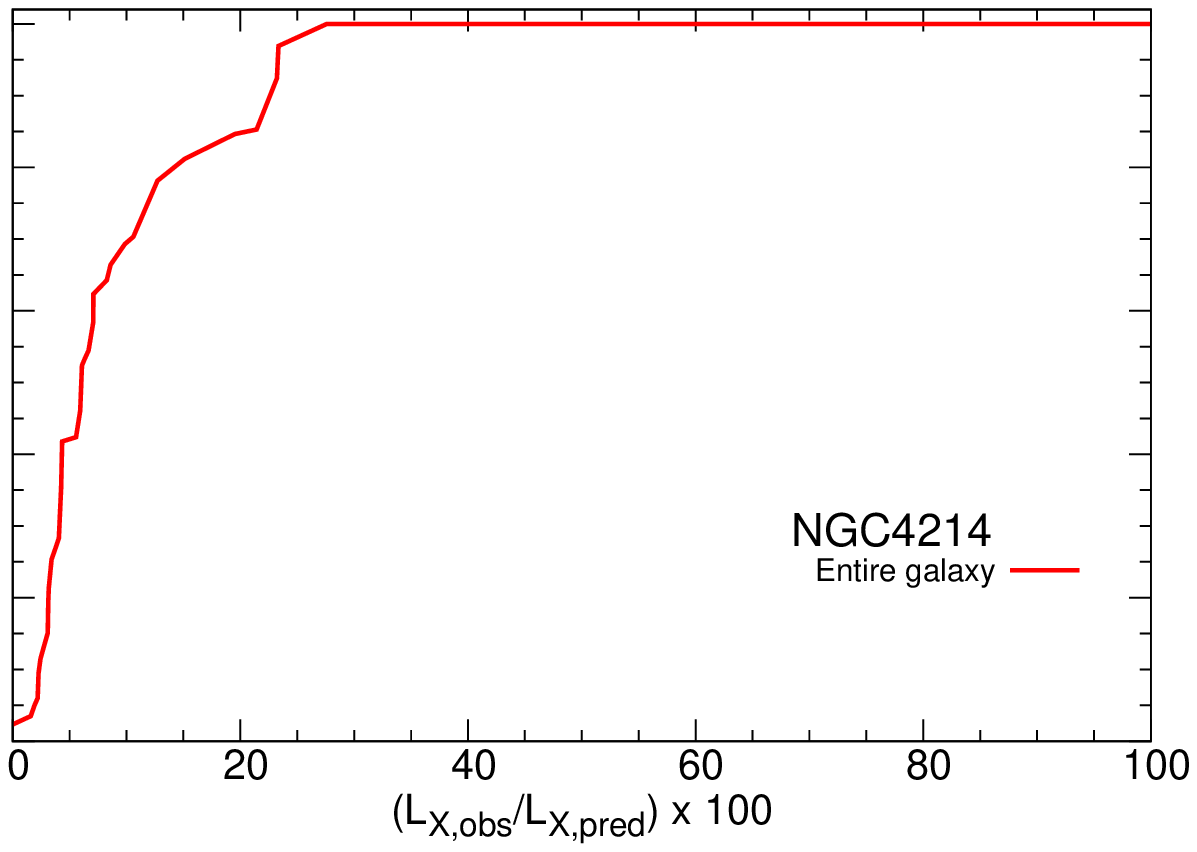}
\hspace{-0.3cm}
\includegraphics[width=6cm]{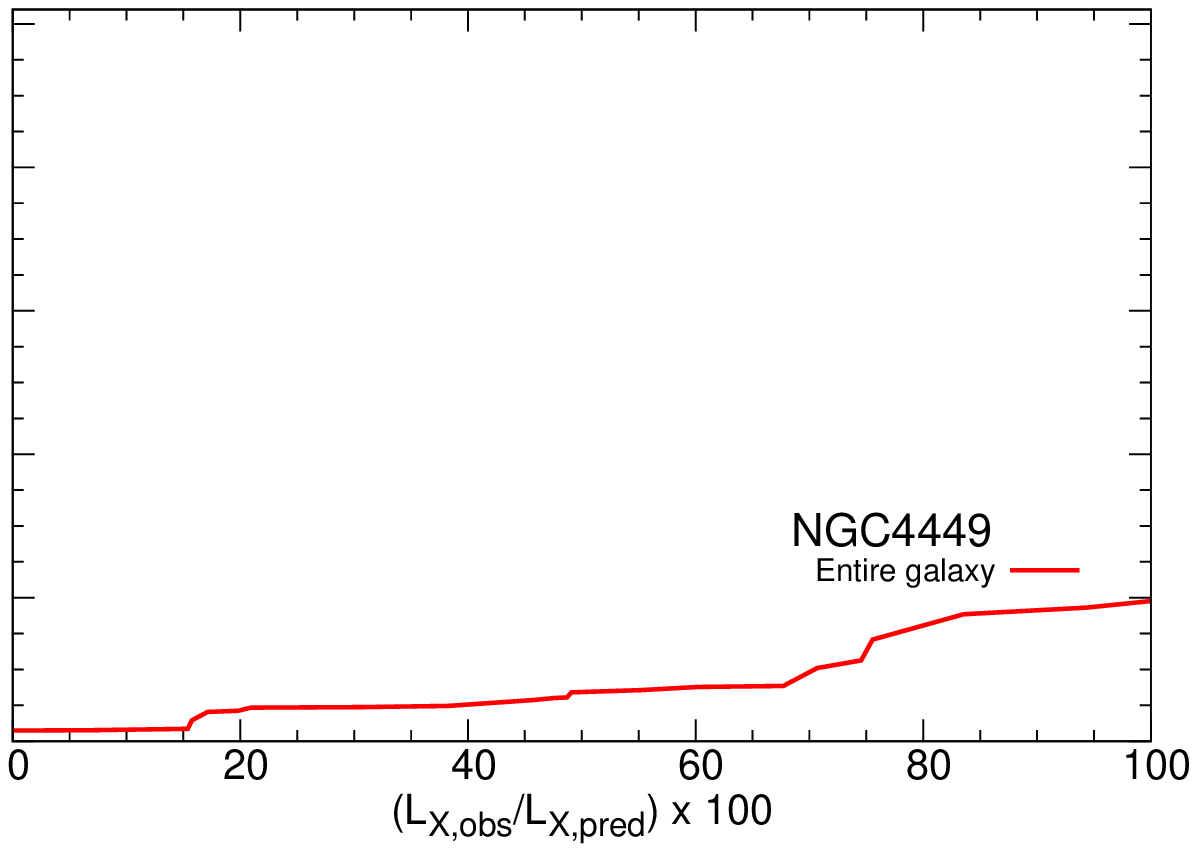}
}
\caption{The fraction of K-band luminosity of the galaxy contained in the cells, in which  the ratio of observed to predicted X-ray luminosities is smaller than a given value. The meaning of this plot is that it shows the fraction of the stellar mass of the galaxy where the contribution of the supersoft sources to the type Ia supernova rate, expressed in per cent,  is smaller than the value shown on the \textit{x}-axis.}  
\label{fig:snfraction}
\end{figure*}

\subsection{Specific frequencies of  supersoft sources}

We also compared the numbers of resolved supersoft sources  detected in bulges and disks and  in spiral arms and in interarm regions.  Due to their small number  in each particular galaxy, we combined data for all nine spiral galaxies. Counting only sources with $L_X\ge10^{36} \ \mathrm{erg \ s^{-1}} $ we found that $16$ of them  are located  in bulges and $51$ in disks.  Taking into account the combined K-band luminosity of bulges ($1.1\cdot10^{11} \ \mathrm{L_{K,\odot}}$) and disks ($2.0\cdot10^{11} \ \mathrm{L_{K,\odot}}$) we derived the specific frequency of the supersoft sources of $\approx1.4$ in bulges and $\approx2.6$ per $10^{10} \ \mathrm{L_{K,\odot}}$ in disks of spiral galaxies. As expected, but probably never confirmed quantitatively before, the specific frequency of supersoft sources in disks exceeds by factor of about $ 2$ that of bulges.

For four grand design spiral galaxies we performed a similar investigation comparing  spiral arms and interarm regions. There were $12$ sources in spiral arms and  $7$  in interarm regions, whose K-band luminosities were   $3.9\cdot10^{10} \ \mathrm{L_{K,\odot}}$ and $1.8\cdot10^{10} \ \mathrm{L_{K,\odot}}$ respectively.  The specific frequencies of supersoft sources are close to each other:  $\approx3.1$ and $\approx 3.9$ per $10^{10} \ \mathrm{L_{K,\odot}}$, respectively. 

As a caveat we note that in calculating statistics of resolved supersoft sources we combined the data for different galaxies having different sensitivity limits (Table 1). In order to reduce this effect we counted only sources with $L_X\ge10^{36} \ \mathrm{erg \ s^{-1}} $, however we made no attempt to perform an accurate incompleteness correction. This may affect somewhat the particular specific frequency values  but would not change results of  the comparison of different structural components of galaxies.

\section{Role of supersoft sources as progenitors of type Ia supernovae}

The observed  $ L_X/L_K $ ratios allow us to constrain the role of supersoft sources as progenitors of SNe Ia. Although unlike the case of early-type galaxies, in late-type galaxies these constraints do not translate into global upper limits on the single degenerate scenario, they are important for understanding relative roles of different evolutionary channels.  We therefore compare observed $ L_X/L_K $ ratios with the ones predicted in the single degenerate scenario. 

In computing the predicted values we assumed that entire process of the mass accumulation by the white dwarfs proceeds  in the regime with the nominal X-ray  radiation efficiency, i.e. that the accreted material has solar composition and the photospheric radius of the nuclear burning hydrogen shell is close to the white dwarf radius.  Particular calculations were conducted using approach and parameters of \citet{nature}, namely the initial white dwarf mass of $ 1.2 \ \mathrm{M_{\odot}} $ and an accretion rate of $ 10^{-7} \ \mathrm{M_{\odot}/yr} $.  
The SN Ia rate was assumed to be proportional to the K-band luminosity with the scale according to  \citet{mannucci05}: 
$\dot{N}_{\mathrm{SNIa}}/L_K = 3.5 \cdot 10^{-4} \ \mathrm{yr^{-1} \ per} \ 10^{10} \ \mathrm{L_{K,\odot}}  $ and 
$ \dot{N}_{\mathrm{SNIa}}/L_K = 8.8 \cdot 10^{-4} \ \mathrm{yr^{-1} \ per} \ 10^{10} \ \mathrm{L_{K,\odot}} $ for the bulges and the disks of spiral galaxies
and $ \dot{N}_{\mathrm{SNIa}}/L_K = 3.3 \cdot 10^{-3} \ \mathrm{yr^{-1} \ per} \ 10^{10} \ \mathrm{L_{K,\odot}} $ for irregular galaxies. 

\begin{figure*}
\hbox{
\includegraphics[width=8.5cm]{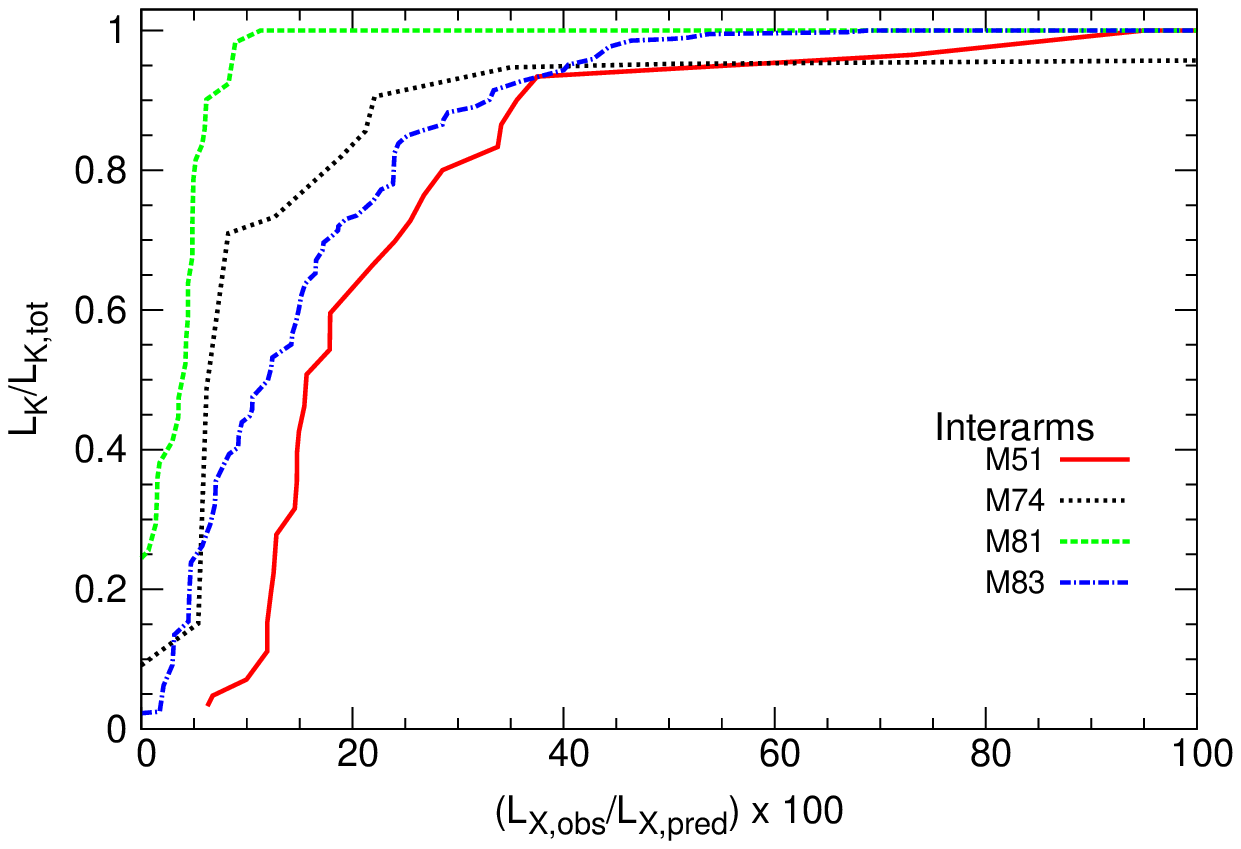}
\hspace{0.3cm}
\includegraphics[width=8.5cm]{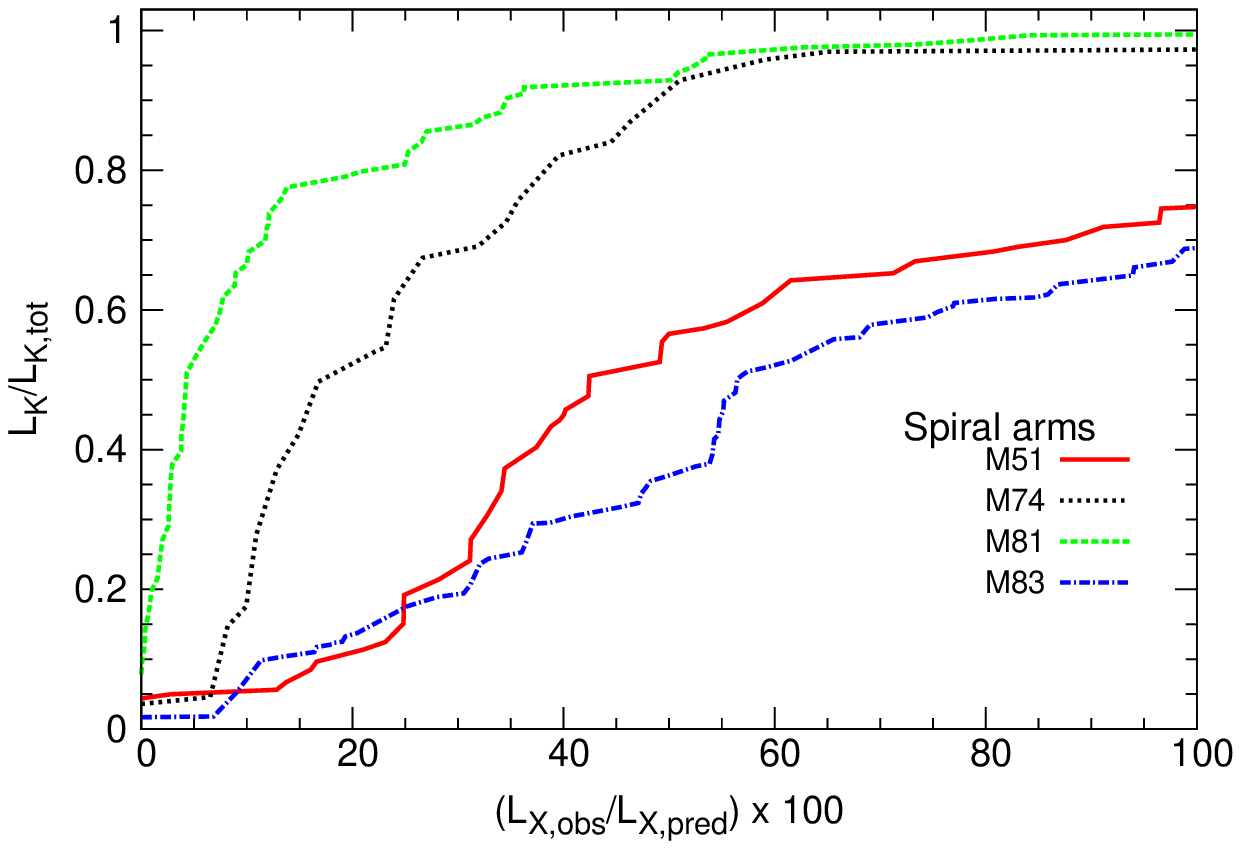}
}
\caption{Same as Fig. \ref{fig:snfraction} but for interarm regions (left panel) and spiral arm regions (right) for the four grand design spiral arm galaxies in our sample.}  
\label{fig:snfraction_arm}
\end{figure*}

\subsection{Galaxy-averaged values}

In comparing predicted luminosities with observations  it is important to take into account strong and spatially varying absorption commonly present in late-type galaxies. In order to accomplish this we divided galaxies in sufficiently small cells. In choosing the cell size we required that each contained  at least $30$ X-ray counts in the $ 0.3-0.7$ keV band, necessary for the spatially resolved analysis described below.  Thus, the number of cells per galaxy depends on their surface brightness  and Chandra exposure time and varied from  $28$ in case of NGC3077 to 1065 for  M101.  The typical K-band luminosity within one cell was in the  $10^7-10^8 \ \mathrm{L_{K,\odot}} $ range and  size was between $ \sim 5-30 \arcsec $. 

In each cell the supernova rate was computed from its K-band luminosity  and the X-ray luminosity of nuclear burning white  dwarfs was computed and absorption (intrinsic, determined for each cell from HI maps, and Galactic) was applied.  The luminosities of all cells in a galaxy were summed and divided by its K-band luminosity to give the average predicted $ L_X/L_K $ ratio. The so computed values are presented in the last column of Table 2. By the method of calculation they present galactic average and are to be compared with observed values listed in the column (4), marked $\left( L_X/L_K \right)_{total}$.

Table \ref{tab:lxlk} shows that for majority of galaxies  predicted $ L_X/L_K $ ratios exceed observed values by a factor of $\sim 7-30$.
For these galaxies we can place  upper limits of $\sim 3-15\%$  on the contribution of supersoft sources to the SN Ia rate. For few galaxies,  M94, NGC 2403 and NGC3077, the difference is $\sim 2$ times. In the case of NGC4449  predicted X-ray luminosity is $\sim 2$ times smaller than the observed value.  For these galaxies no meaningful upper limits can be placed.
We note that the latter galaxy has largest $N_{\mathrm{H}}$ and $ L_X/L_K $ ratio. It is further discussed in the Sect. \ref{sec:nhsfr}.

\subsection{Spatially resolved analysis}

Using the results of this calculation it is also possible to compare prediction of the model with observation  in a more detailed, spatially resolved manner, computing the ratio of the observed luminosity to the predicted value in each individual cell. To present results of this analysis  we plot in the Fig. \ref{fig:snfraction}  the fraction of K-band luminosity of the galaxy contained in the cells, in which  the ratio of observed to predicted X-ray luminosity is smaller than a given value. The meaning of this plot is that it shows the fraction of the stellar mass of the galaxy where the contribution of the supersoft sources to the type Ia supernova rate is smaller than the given value.

The plot reveals a large variations  between galaxies in our sample. In several of them, M74, M81, M95 and NGC3184,  supersoft sources cannot contribute more than $\sim 5-10\%$ to the type Ia supernova rate in more than $\sim 80-90\%$ of their mass. The other end of the range is represented by NGC2403 and NGC4449, in which supersoft sources could, in principle, be progenitors of a large fraction of type Ia supernovae without contradicting X-ray observations.

\subsection{Spiral arms and interarm regions}

We applied similar approach to analyze spiral arms and interarm regions in the four grand design spiral galaxies. 
In computing the SN Ia rates we note that scale factors from  \citet{mannucci05} refer to entire galaxies, therefore it may be inaccurate to use them in the analysis of arm and interarm regions. Instead, we use results of \citet{sullivan06} who decomposed the SN Ia rate into a mass related and a star-formation related components. Namely, we assumed scale factors  of 
$ 5.3 \cdot 10^{-14} \ \mathrm{SNe \ yr^{-1} \ M_{\odot}^{-1}}  $ and  
$ 3.9 \cdot 10^{-4} \ \mathrm{SNe \ yr^{-1} \ (M_{\odot} \ yr^{-1})}  $. 
The  stellar mass in cells was determined from their K-band luminosity  assuming a mass-to-light ratio of $M_{\star}/L_{K}=0.8 $. The star-formation rate was computed based on \textit{Spitzer} far-infrared images as described in Sect. \ref{sec:infrared}. 
The resulting upper limits are plotted  in Fig. \ref{fig:snfraction_arm} using the same approach as in Fig. \ref{fig:snfraction}. The plot shows that upper limits for interarm regions are generally tighter than for the disks as a whole. Overall, in more than $\sim 80\%$ of mass the contribution of supersoft sources to the SN Ia rate  is less than $\sim 5-30\%$. In the spiral arms, on the contrary, no constraining upper limits can be obtained in three out of four galaxies -- the supersoft sources can, in principle, be progenitors of a large fraction of SN Ia. 

A number of studies \citep[e.g.][]{bildsten05,mannucci06} indicate that SNe Ia originate from two sub-populations, often designated as prompt and delayed populations. The prompt population is believed to dominate at ages $ \lesssim 10^8 $ years, therefore this component is presumably associated with star-forming regions of galaxies, whereas the delayed population is producing SNe Ia in older stellar populations. According to the results of our spatially resolved analysis we find that a model, in which the major fraction of the prompt population originates from supersoft sources, does not contradict our X-ray data.

\begin{figure*}
\hbox{
\includegraphics[width=8.5cm ]{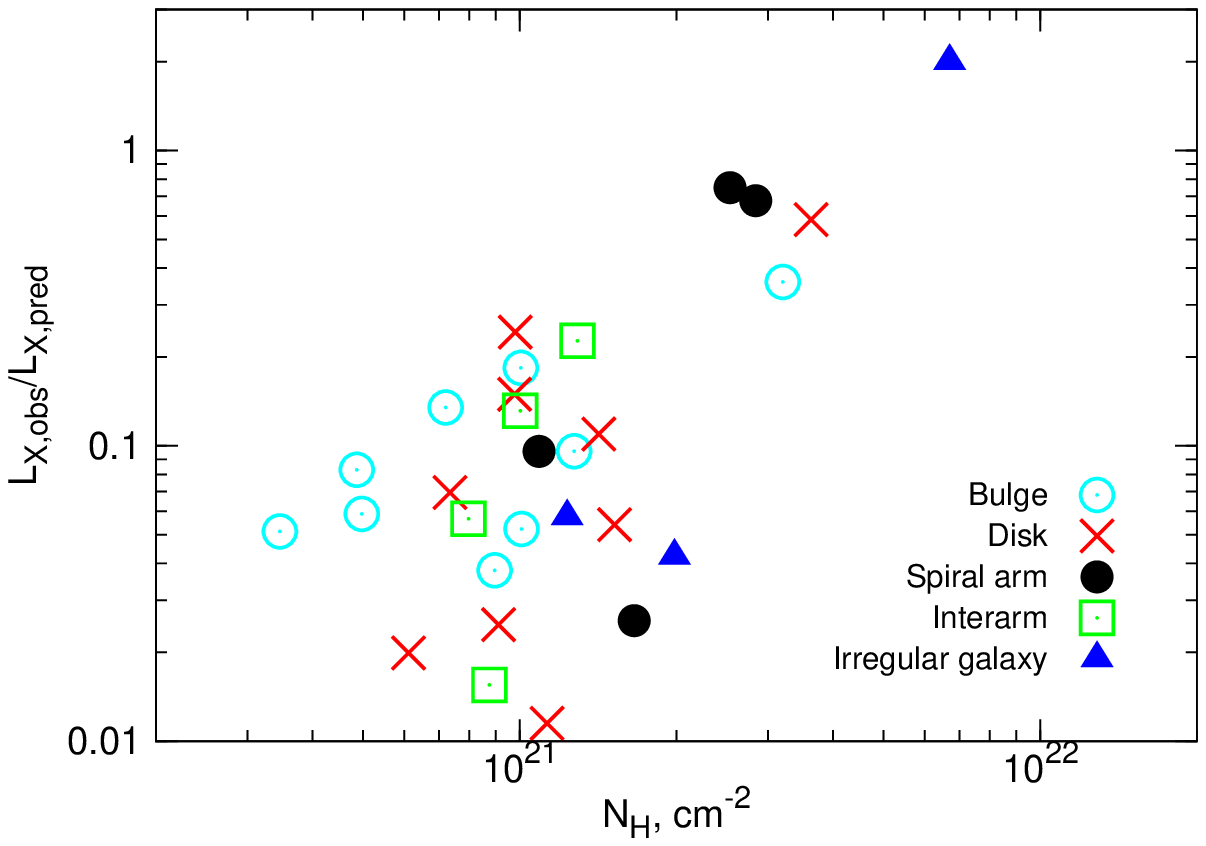}
\hspace{0.3cm}
\includegraphics[width=8.5cm]{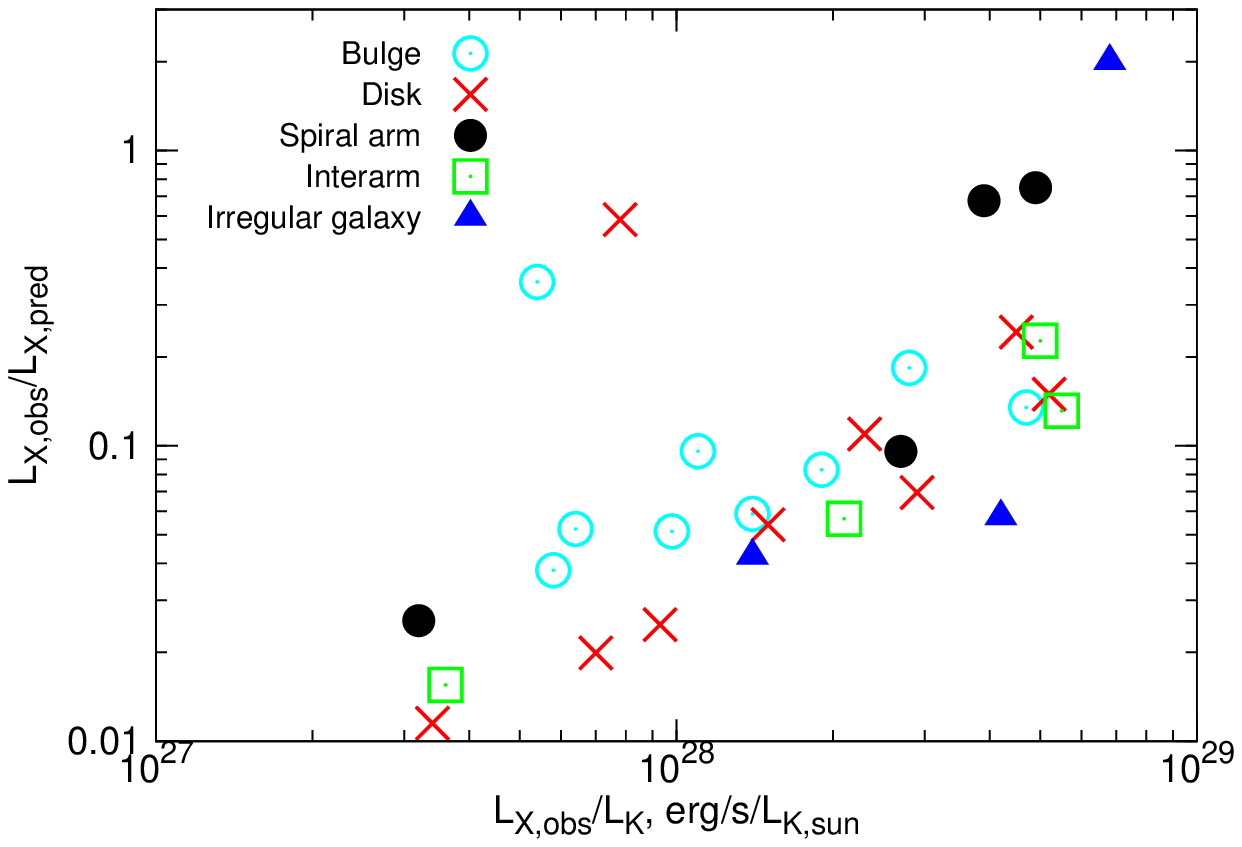}
}
\caption{Ratio of observed  luminosity in the $0.3-0.7$ keV band  to the luminosity of supersoft sources predicted in the single degenerate scenario,  for entire galaxies and their structural components, versus average $N_{\mathrm{H}}$ and $ L_X/L_K $ ratio.}
\label{fig:lxr1}
\end{figure*}

\subsection{Dependence of upper limits on absorption and star-formation rate}
\label{sec:nhsfr}

The obtained upper limits vary significantly between  galaxies, from a few per cent level to unconstraining values of tens of per cent. One may ask a question, whether these variations are a result of different levels of contamination and absorption in different galaxies or do indeed reflect varying contribution of supersoft sources. In an attempt to answer this question we plot in Fig. \ref{fig:lxr1} the ratio of the observed X-ray luminosity to the luminosity of supersoft sources predicted in the single degenerate scenario, $L_X^{obs}/L_X^{pred}$, versus the $N_{\mathrm{H}} $ value and $ L_X/L_K $ ratio. The predicted luminosity was computed by the method described earlier in this section. The plots show clear correlations -- the observed-to-predicted luminosity ratio increases with the column density and with specific luminosity of unresolved X-ray emission. This may indicate that variations in the upper limits are caused by  variations in absorption and in  the level of contamination.

\begin{figure}
\hbox{
\includegraphics[width=8.5cm]{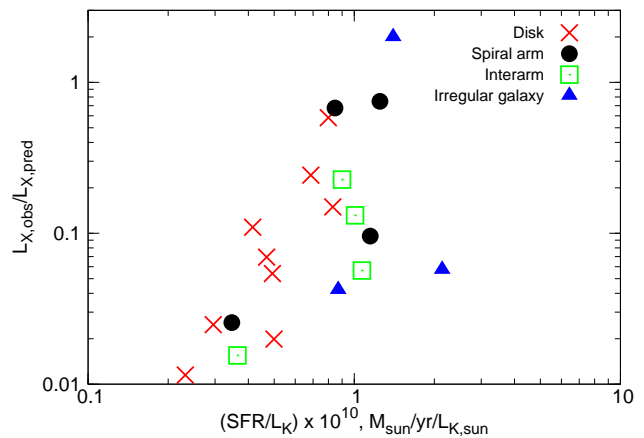}
}
\caption{Ratio of observed  luminosity in the 0.3--0.7 keV band  to the luminosity of supersoft sources predicted in the single degenerate scenario,  for entire galaxies and their structural components,  versus star-formation rate.}
\label{fig:lxr2}
\end{figure}

On the other hand, in a picture, which is not entirely implausible, the  role of supersoft sources as  SN Ia progenitors  may increase in star-forming galaxies.  Therefore one may expect  a positive correlation of the luminosity of supersoft sources with the star-formation rate.   To investigate this, we plot in Fig. \ref{fig:lxr2} the observed-to-predicted luminosity ratio versus specific star-formation rate. The plot does show some evidence of such a correlation, in an apparent contradiction to the tentative conclusion made above.  
It should be taken into account however, that  star-forming galaxies tend to have larger $N_{\mathrm{H}} $ and brighter levels of the diffuse emission, which may produce secondary correlations. 

\section{Conclusions}

The aim of this study was to measure soft band $ L_X/L_K $ ratios in late-type galaxies. Although these ratios are of interest on their own, our primary goal was to derive constraints on X-ray  emission from nuclear burning white dwarfs and to constrain  the role   of supersoft sources as progenitors of SNe Ia. To this end we used an extensive set of archival \textit{Chandra} observations of a sample of twelve late-type galaxies, which included nine spiral and three irregular galaxies. In selecting our sample we demanded a source detection sensitivity of at least $5\cdot 10^{36} \ \mathrm{erg \ s^{-1}}$ allowing the removal of bright X-ray binaries. We also required the existence of publicly available HI maps in order  to measure the intrinsic absorption.

The observed  $0.3-0.7$ keV band $L_X/L_K$ ratios  range from  $5.4\cdot10^{27}$ to   $6.8\cdot10^{28} \ \mathrm{erg \ s^{-1} \ L_{K,\odot}^{-1}}$, significantly exceeding those in early-type galaxies. Based on the spectral analysis we tentatively suggest  that the primary reason for this difference is the soft X-ray emission from hot ionized  ISM  with the temperature of $kT\sim 0.2-0.4$ keV.   X-ray emission from  young stars in star-forming regions may also contribute. On a more detailed level we determine $ L_X/L_K $ ratios for bulges and disks of spiral galaxies separately  and for majority of them found a factor of $\sim 2-4$ difference. Surprisingly,  the sign of this difference was different in different galaxies, indicating the complexity of distribution of various X-ray emitting components in spiral  galaxies. Furthermore, in four  grand design spiral galaxies  we separated spiral arms and interarm regions and found that their $L_X/L_K$ ratios do not differ by more than $\sim30\%$. 

We investigated the role of supersoft sources as progenitors of SNe Ia. As a measure of their contribution we calculated the ratio of the observed soft X-ray luminosity to the luminosity predicted by the single degenerate scenario. The latter was computed assuming the nominal X-ray radiation efficiency of nuclear burning white dwarfs, as expected  in the standard picture of supersoft sources \citep{kahabka97}. We find that for majority of galaxies the predicted luminosity exceeds observed values  by a factor of $\sim 7-30$. Taken at the face value, these numbers  imply upper limit of $\sim 3-15\%$  on the contribution of supersoft sources to the SN Ia rate. For a few galaxies observed and predicted luminosities are comparable, therefore  no constraining upper limits can be placed, including NGC4449, where predicted luminosity is smaller than the observed one.  The spatially resolved analysis yielded similar results. 
Correlations of the upper limits with $N_{\mathrm{H}} $ and $ L_X/L_K $ ratios suggest that large scatter in upper limit may be a result of different level of contamination and absorption in different galaxies.   

It is important to realize that for the late-type galaxies, upper limits on the contribution of supersoft sources to SN Ia rate do not translate into upper limits on the single degenerate scenario in general. In young stellar environment several possibilities  may exist, that X-ray emission emission from nuclear burning white dwarfs is significantly suppressed  \citep[e.g.][]{hachisu96,iben94}. Note that these possibilities are are not expected to play significant role in early-type galaxies considered in \citet{nature}, due to the old age of their stellar populations. 

Finally, we investigated the specific frequency of bright supersoft sources in bulges and disks of spiral galaxies and found significant difference, the supersoft sources being factor of $\sim 2$ times more frequent (per unit K-band luminosity) in the disks than in the bulges.

\bigskip
\begin{small}

\noindent
\textit{Acknowledgements.}
We thank the anonymous referee for his/her useful comments. \'A. Bogd\'an is grateful to Stefano Mineo for helpful discussions about measurements of star formation rates. This research has made use of \textit{Chandra} archival data provided by the \textit{Chandra} X-ray Center. The publication makes use of software provided by the \textit{Chandra} X-ray Center (CXC) in the application package \begin{small}CIAO\end{small}. The \textit{Spitzer Space Telescope} is operated by the Jet Propulsion Laboratory, California Institute of Technology, under contract with the National Aeronautics and Space Administration. This publication makes use of data products from the Two Micron All Sky Survey, which is a joint project of the University of Massachusetts and the Infrared Processing and Analysis Center/California Institute of Technology, funded by the National Aeronautics and Space Administration and the National Science Foundation. This work makes use of THINGS, ``The HI Nearby Galaxy Survey''.
\end{small}


\begin{thebibliography}{}
\bibitem[\protect\citeauthoryear{Bavouzet et al.}
 {Bavouzet et al.}{2008}]{bavouzet08}
  Bavouzet, N., Dole, H., Le Floc'h, E., Caputi, K. I., Lagache, G., Kochanek, C. S., 2008, A\&A, 479, 83
\bibitem[\protect\citeauthoryear{Bell}
 {Bell}{2003}]{bell03}
  Bell, E. F., 2003, ApJ, 586, 794
\bibitem[\protect\citeauthoryear{Bogd\'an \& Gilfanov}
 {Bogd\'an \& Gilfanov}{2008}]{bogdan08}
  Bogd\'an, \'A. \& Gilfanov, M., 2008, MNRAS, 388, 56
\bibitem[\protect\citeauthoryear{Bogd\'an \& Gilfanov}
 {Bogd\'an \& Gilfanov}{2010}]{bogdan10}
  Bogd\'an, \'A. \& Gilfanov, M., 2010, A\&A, 512, 16
\bibitem[\protect\citeauthoryear{Dickey \& Lockman}
 {Dickey \& Lockman}{1990}]{dickey90}
  Dickey, J. M., Lockman, F. J., 1990, ARA\&A, 28, 215
\bibitem[\protect\citeauthoryear{Di Stefano \& Kong}
 {Di Stefano \& Kong}{2003}]{distefano03}
  Di Stefano, R. \& Kong, A. K. H., 2003, ApJ, 592 884
\bibitem[\protect\citeauthoryear{Di Stefano}
 {Di Stefano}{2010}]{distefano10}
 Di Stefano, R., 2010, ApJ, 712, 728
\bibitem[\protect\citeauthoryear{Freedman et al.}
 {Freedman et al.}{2001}]{freedman01}
  Freedman, W. L., et al., 2001, ApJ, 553, 47
\bibitem[\protect\citeauthoryear{Gilfanov}
 {Gilfanov}{2004}]{gilfanov04}
 Gilfanov, M., 2004, MNRAS, 349, 146 
\bibitem[\protect\citeauthoryear{Gilfanov \& Bogd\'an}
 {Gilfanov \& Bogd\'an}{2010}]{nature}
 Gilfanov, M. \& Bogd\'an, \'A., 2010, Nature, 463, 924
\bibitem[\protect\citeauthoryear{Grimm et al.}
 {Grimm et al.}{2003}]{grimm03}
  Grimm, H.-J., Gilfanov, M. \& Sunyaev, R., MNRAS, 339, 793
\bibitem[\protect\citeauthoryear{Hachisu et al.}
 {Hachisu et al.}{1996}]{hachisu96}
 Hachisu, I., Kato, M. \& Nomoto, K., 1996, ApJ, 470, 97
\bibitem[\protect\citeauthoryear{Hachisu et al.}
 {Hachisu et al.}{1999}]{hachisu99}
 Hachisu, I., Kato, M. \& Nomoto, K., 1999, ApJ, 522, 487
\bibitem[\protect\citeauthoryear{Iben \& Tutukov}
 {Iben \& Tutukov}{1994}]{iben94}
 Iben, I. Jr. \& Tutukov, A. V. 1994, ApJ, 431, 264
\bibitem[\protect\citeauthoryear{Irwin et al.}
 {Irwin et al.}{2003}]{irwin03}
 Irwin, J. A., Athney, A. E. \& Bregman, J. N. 2003, ApJ, 587, 356
\bibitem[\protect\citeauthoryear{Jarrett et al.}
 {Jarrett et al.}{2003}]{jarrett03}
 Jarrett, T. H., Chester, T., Cutri, R., et al., 2003, AJ, 125, 525
\bibitem[\protect\citeauthoryear{Kahabka \& van den Heuvel}
 {Kahabka \& van den Heuvel}{1997}]{kahabka97}
  Kahabka, P. \& van den Heuvel, E. P. J., 1997, ARA\&A, 35, 69
\bibitem[\protect\citeauthoryear{Karachentsev et al.}
 {Karachentsev et al.}{2004}]{karachentsev04}
 Karachentsev, I. D., Karachentseva, V. E., Huchtmeier, W. K. \& Makarov, D. I., 2004, AJ, 127, 2031
\bibitem[\protect\citeauthoryear{Leonard et al.}
 {Leonard et al.}{2004}]{leonard02}
  Leonard, D. C., et al., 2002, AJ, 124, 2490
\bibitem[\protect\citeauthoryear{Mannucci et al.}
 {Mannucci et al.}{2005}]{mannucci05}
  Mannucci, F., Della Valle, M., Panagia, N., Cappellaro, E., Cresci, G., Maiolino, R., Petrosian, A. \& Turatto, M., 2005, A\&A, 433, 807
\bibitem[\protect\citeauthoryear{Mannucci et al.}
 {Mannucci et al.}{2006}]{mannucci06}
  Mannucci, F., Della Valle, M. \& Panagia, N., 2006, MNRAS, 370, 773
\bibitem[\protect\citeauthoryear{Nomoto et al.}
 {Nomoto et al.}{2007}]{nomoto07}
  Nomoto, K., Saio, H., Kato, M. \& Hachisu, I., 2007, ApJ, 663, 1269
\bibitem[\protect\citeauthoryear{Perlmutter et al.}
 {Perlmutter et al.}{1999}]{perlmutter99}
  Perlmutter, S., Aldering, G., Goldhaber, G., et al., 1999, ApJ, 517, 565
\bibitem[\protect\citeauthoryear{Riess et al.}
 {Riess et al.}{1998}]{riess98}
  Riess, A. G., Filippenko, A. V., Challis, P., et al., 1998, AJ, 116, 1009
\bibitem[\protect\citeauthoryear{Sazonov et al.}
 {Sazonov et al.}{2006}]{sazonov06}
 Sazonov, S., Revnivtsev, M., Gilfanov, M., Churazov, E. \& Sunyaev, R., 2006, A\&A, 450, 117
\bibitem[\protect\citeauthoryear{Scannapieco \& Bildsten}
 {Scannapieco \& Bildsten}{2005}]{bildsten05}
  Scannapieco, E. \& Bildsten, L., 2005, ApJ, 629, 85
\bibitem[\protect\citeauthoryear{Shtykovskiy \& Gilfanov}
 {Shtykovskiy \& Gilfanov}{2005}]{shtykovskiy05}
  Shtykovskiy, P. \& Gilfanov, M., 2005, MNRAS, 362, 879
\bibitem[\protect\citeauthoryear{Sullivan et al.}
 {Sullivan et al.}{2006}]{sullivan06}
  Sullivan, M., et al., 2006, ApJ, 648, 868
\bibitem[\protect\citeauthoryear{Walter et al.}
 {Walter et al.}{2008}]{walter08}
  Walter, F., Brinks, E., de Blok, W. J. G., Bigiel, F., Kennicutt, R. C., Thornley, M. D. \& Leroy, A., 2008, AJ, 136, 2563
\end{thebibliography}
\end{document}